\documentclass[prd,notitlepage,longbibliography,nofootinbib,superscriptaddress,onecolumn,preprintnumbers]{revtex4-2}
\usepackage[utf8]{inputenc}

\usepackage{bm}
\usepackage{comment} 
\usepackage[colorlinks=true,urlcolor=blue,anchorcolor=black,citecolor=blue,linkcolor=red,filecolor=black,menucolor=black,pagecolor=black,linktocpage=true,pdfproducer=medialab,pdfa=true]{hyperref}
\usepackage{graphicx}
\usepackage{amsmath,latexsym,amssymb,mathrsfs,ascmac,mathtools}
\usepackage{multirow}
\usepackage{braket}
\usepackage{placeins}
\allowdisplaybreaks
\begin{document}

\title{Analytical Study of Deflection Angle and Time Delay in Kerr Spacetime with Modified Propagation}

\author{Takamasa Kanai}
\email{kanai@kochi-ct.ac.jp}

\affiliation{Department of Social Design Engineering,
National Institute of Technology (KOSEN), Kochi College,
200-1 Monobe Otsu, Nankoku, Kochi, 783-8508, Japan}

\begin{abstract}
We investigate gravitational lensing in Kerr spacetime in the presence of a modified photon propagation law arising from higher-curvature effective field theory corrections. Adopting a deformed dispersion relation, we analytically derive the deflection angle and the propagation time delay for null trajectories in a rotating background.

We show that the modified propagation leads to explicit and calculable deviations from the standard Kerr predictions, affecting both the bending angle and the time delay. These corrections exhibit a nontrivial dependence on the black hole spin and are expected to become more relevant in the strong-field regime. In particular, near the photon region—where the spin-dependent geometry influences photon trajectories—such effects may play a role in shaping observable propagation features.

Our results establish a concrete and systematic framework to quantify deviations from standard photon propagation in gravitational lensing. They further indicate that observables such as relativistic image separations and time delays provide a potential avenue to probe ultraviolet corrections to gravity in strong-field environments.
\end{abstract}
\maketitle

\section{Introduction}

Gravitational phenomena in the strong-field regime provide a unique window into the fundamental nature of gravity beyond the weak-field approximation. In particular, null geodesics propagating near compact objects exhibit universal features governed by unstable photon orbits, which are described geometrically by photon surfaces \cite{Virbhadra:1999nm,Claudel:2000yi}. These structures underlie a wide range of observables, including black hole shadows  \cite{Falcke:1999pj,EventHorizonTelescope:2019dse,EventHorizonTelescope:2022wkp,Vagnozzi:2022moj} and gravitational lensing \cite{Virbhadra:1999nm,Bozza:2001xd,Bozza:2002zj,Gibbons:2008rj}. A coordinate-invariant formulation of the strong deflection coefficient a, expressed in terms of local geometric quantities evaluated along the unstable circular photon orbit, has been developed in Refs.~\cite{Igata:2025taz,Igata:2025hpy,Igata:2026hzb}.

Among these, strong gravitational lensing offers a particularly direct probe of the near-horizon geometry. In the strong deflection limit, the bending angle diverges logarithmically as light rays approach the photon sphere. This behavior is captured by the formalism developed by Bozza \cite{Bozza:2002zj,Bozza:2002af,Bozza:2003cp}, where the deflection angle is expressed in terms of a universal logarithmic divergence and a finite regular contribution. Importantly, the coefficients appearing in this expansion are determined entirely by local geometric quantities evaluated at the photon sphere.

This locality makes strong lensing observables especially sensitive to deviations from general relativity. Within the framework of effective field theory (EFT) \cite{Weinberg:1978kz,Donoghue:1993eb,Donoghue:1994dn,Burgess:2003jk}, higher-curvature corrections generically modify both the background geometry and the propagation of null rays. Consequently, key quantities such as the photon sphere radius and the critical impact parameter receive corrections controlled by EFT couplings, which are encoded in the coefficients of the strong deflection expansion.

Despite this connection, a systematic understanding of how EFT corrections are reflected in strong lensing observables remains incomplete. In particular, it is not yet clear to what extent the coefficients of the deflection angle can be used to extract or constrain the underlying EFT parameters in a model-independent way.

In general relativity, photon propagation follows null geodesics of the spacetime metric, and the causal structure is determined by null hypersurfaces corresponding to the characteristics of the Maxwell equations. In contrast, in EFTs with higher-curvature corrections, higher-derivative interactions between curvature and gauge fields modify the propagation of electromagnetic waves. As a result, the characteristic surfaces no longer coincide with the null cones of the background metric (see, e.g., Refs.~\cite{Scharnhorst:1990sr,Barton:1989dq,Barton:1992pq,Latorre:1994cv,Dittrich:1998fy,DeLorenci:2000yh} in flat spacetime and Refs.~\cite{Drummond:1979pp,Daniels:1993yi,Shore:1995fz,Daniels:1995yw,Shore:2007um,Cho:1997vg,Izumi:2014loa,Reall:2014pwa,Allahyari:2019jqz,Cao:2021sty,Reall:2021voz,Davies:2021frz,Fu:2025oxr} in curved spacetimes). Instead, photon propagation is governed by effective characteristic surfaces, which can differ for each polarization mode, leading to gravitational birefringence.

In this paper, we address this issue by performing a systematic analysis of strong gravitational lensing within the framework of effective field theory in general static and spherically symmetric spacetimes. Using the strong deflection limit formalism, we derive the general structure of the deflection angle and explicitly compute both the divergent and regular contributions.

In addition, we discuss the behavior in regimes corresponding to both small and large black hole angular momentum, including the extremal limit. For concreteness, we carry out our analysis by fixing the mass parameter $M$ to a specific value; however, this choice does not affect the generality of our results.

The main novelty of this work lies in incorporating modified photon propagation beyond standard null geodesics into the strong deflection framework and systematically analyzing its impact on gravitational lensing observables. While previous studies have primarily focused on modifications of the background geometry, we instead consider the effects of higher-curvature interactions on the photon dynamics itself, leading to a deformed propagation law. This allows us to derive, for the first time, analytic expressions for the deflection angle and time delay that explicitly encode both rotational effects and effective field theory corrections.

In particular, we demonstrate how the coefficients of the strong deflection expansion are modified by the interplay between black hole spin and higher-curvature couplings, thereby providing a direct link between EFT parameters and observable quantities. This establishes a new framework in which strong gravitational lensing can be used as a probe of modified photon propagation and, more generally, of ultraviolet corrections to gravity.

Our goal is to provide a clear and systematic characterization of how the coefficients in the strong deflection expansion encode the properties of the photon sphere and how they are modified by higher-curvature corrections. By deriving explicit analytic expressions that incorporate both rotational effects and modified photon propagation, we clarify the extent to which strong lensing observables can be used to extract or constrain EFT couplings in a model-independent manner.

In Sec.~\ref{sec:photon_surface}, we review the concept of photon surfaces and their key properties in static and stationary spacetimes. In Sec.~\ref{sec.kerr}, we investigate modifications to photon propagation in Kerr spacetime. In Sec.~\ref{sec:strong lens}, we analyze gravitational lensing in the strong-field regime. In Sec.~\ref{sec:time delay}, we study the time delay in the strong-field regime. Finally, in Sec.~\ref{sec:conclusion}, we summarize our results and discuss future prospects.

In this paper, we set the Newton constant and the speed of light equal to unity.

\section{Photon Surface}
\label{sec:photon_surface}

We briefly review the notion of a photon surface following Ref.~\cite{Claudel:2000yi}. Photon surfaces play a central role in strong gravitational fields, as they govern the behavior of null geodesics and are directly related to observables such as photon rings and strong gravitational lensing.

\subsection{Definition and geometric characterization}

A \textit{photon surface} $\mathcal{S}$ is a timelike hypersurface in a spacetime $(M, g_{ab})$ such that any null geodesic initially tangent to $\mathcal{S}$ remains confined to it \cite{Claudel:2000yi}. This definition generalizes the notion of a photon sphere without assuming symmetry.

Geometrically, let $n^a$ be the unit spacelike normal to $\mathcal{S}$. The induced metric and extrinsic curvature are defined as
\begin{align}
    h_{ab} &= g_{ab} - n_a n_b, \\
    K_{ab} &= h_a{}^c \nabla_c n_b.
\end{align}
The photon surface condition is equivalent to requiring that $\mathcal{S}$ be \textit{umbilical},
\begin{equation}
    K_{ab} \propto h_{ab},
\end{equation}
which provides a simple geometric characterization.

In spherically symmetric spacetimes, the photon surface reduces to the familiar photon sphere (e.g., $r=3M$ in Schwarzschild). In contrast, in rotating spacetimes such as Kerr, no global photon surface exists; instead, null geodesics form an extended photon region depending on their conserved quantities \cite{Teo:2003ltt,Perlick:2004tq,Grenzebach:2014fha}.

\subsection{Remarks on uniqueness}

Uniqueness theorems suggest that asymptotically flat vacuum spacetimes admitting a static photon surface are highly constrained. In particular, under suitable assumptions, such spacetimes are expected to be isometric to the Schwarzschild solution \cite{Cederbaum:2014gva,Cederbaum:2015fra}. Perturbative analyses further support this expectation, indicating that nontrivial deformations of photon surfaces are not allowed in vacuum without introducing finely tuned matter configurations \cite{Yoshino:2016kgi}.

\section{Review of the photon-Weyl-tensor coupling in a Kerr spacetime}
\label{sec.kerr}

Gravitational lensing by compact objects is a well-established probe of
strong-field gravity and of the matter distribution surrounding black
holes~\cite{Einstein:1936llh,Virbhadra:1999nm,Bozza:2001xd,Bozza:2002zj,Gibbons:2008rj}. Beyond the
geometry of the background spacetime, the trajectory of light can also be
affected by non-minimal couplings between the photon and the spacetime
curvature. Couplings of this kind arise naturally from one-loop vacuum
polarization in curved space~\cite{Drummond:1979pp} and have since been
studied with phenomenologically arbitrary coupling strengths, motivated
for instance by the generation of large-scale magnetic fields and by
photon propagation near compact astrophysical
sources~\cite{Ni:1977zz,Solanki:2004az,Dereli:2011mk}. Among the possible
curvature couplings, the interaction between the electromagnetic field
and the \emph{Weyl tensor} is of particular interest because it appears
in holographic descriptions of strongly coupled condensed-matter systems
and modifies the optical properties of charged black
holes~\cite{Ritz:2008kh}. Photon--Weyl couplings have previously been
shown to alter the strong-field deflection angle, the resulting
relativistic images, and the associated time delays for a
\emph{non-rotating} (Schwarzschild) black
hole~\cite{Drummond:1979pp,Chen:2015cpa,Lu:2016gsf}. Since
astrophysical compact objects generically rotate, Chen \emph{et
al.}~\cite{Chen:2016hil} extend this analysis to a Kerr
background, deriving the equation of motion for a Weyl-coupled photon and
the associated effective optical metric that governs its strong-field
lensing. The remainder of this section reviews that derivation, which we
use as the starting point for the calculations below.

\subsection{Equation of motion for a Weyl-coupled photon}

The starting point is the action for an electromagnetic field coupled to
the Weyl tensor $C_{\mu\nu\rho\sigma}$ in a curved background,
\begin{equation}
S=\int d^{4}x\sqrt{-g}\left[\frac{R}{16\pi G}-\frac{1}{4}\Big(F_{\mu\nu}F^{\mu\nu}-4\alpha\, C^{\mu\nu\rho\sigma}F_{\mu\nu}F_{\rho\sigma}\Big)\right],
\label{eq:action}
\end{equation}
where $F_{\mu\nu}$ is the Maxwell field strength and $\alpha$, with
dimension of length squared, measures the strength of the coupling to
\begin{equation}
C_{\mu\nu\rho\sigma}=R_{\mu\nu\rho\sigma}-\big(g_{\mu[\rho}R_{\sigma]\nu}-g_{\nu[\rho}R_{\sigma]\mu}\big)+\tfrac{1}{3}R\,g_{\mu[\rho}g_{\sigma]\nu}.
\label{eq:weyl-def}
\end{equation}
Varying the action~\eqref{eq:action} gives a Weyl-corrected Maxwell
equation,
\begin{equation}
\nabla_{\mu}\!\left(F^{\mu\nu}-4\alpha\,C^{\mu\nu\rho\sigma}F_{\rho\sigma}\right)=0 .
\label{eq:maxwell-mod}
\end{equation}
For a vacuum background such as the Kerr spacetime, higher-curvature corrections to photon propagation arise only through the above $R_{\mu\nu\rho\sigma}F^{\mu\nu}F^{\rho\sigma}$ coupling at leading order in the perturbative expansion. Substituting the background metric into this term and treating it as a source, one finds that there is no nontrivial first-order correction to the photon equation of motion.
From the viewpoint of a derivative expansion, higher-order curvature corrections, such as cubic terms, are expected to contribute at subleading orders compared to the $R_{\mu\nu\rho\sigma}F^{\mu\nu}F^{\rho\sigma}$ term. Therefore, when considering gravitational lensing up to first order in higher-curvature corrections, it is sufficient to adopt the Kerr spacetime as the background geometry and account for modifications to photon propagation induced by this coupling.

The photon equation of motion follows from
Eq.~\eqref{eq:maxwell-mod} in the geometric-optics (eikonal) limit, in
which the photon wavelength is much shorter than the curvature scale of
the background but still much longer than the electron Compton
wavelength. Writing $F_{\mu\nu}=f_{\mu\nu}e^{i\theta}$ for a slowly
varying amplitude $f_{\mu\nu}$ and rapidly varying phase $\theta$, the
wave vector is $k_{\mu}=\partial_{\mu}\theta$, and the Bianchi identity
forces $f_{\mu\nu}=k_{\mu}a_{\nu}-k_{\nu}a_{\mu}$ for a polarization
vector $a_{\mu}$ satisfying $k_{\mu}a^{\mu}=0$. Substituting this ansatz
into Eq.~\eqref{eq:maxwell-mod} yields the equation of motion for the
coupled photon,
\begin{equation}
k_{\mu}k^{\mu}a^{\nu}+8\alpha\,C^{\mu\nu\rho\sigma}k_{\sigma}k_{\mu}a_{\rho}=0 ,
\label{eq:eom-photon}
\end{equation}
which shows explicitly that the Weyl coupling modifies the light cone
felt by the photon relative to the geodesics of the background metric.

\subsection{Kerr background and the Weyl tensor in an orthonormal frame}

The background is the Kerr metric in Boyer-Lindquist coordinates,
\begin{equation}
ds^{2}=-\rho^{2}\frac{\Delta}{\Sigma^{2}}dt^{2}+\frac{\rho^{2}}{\Delta}dr^{2}+\rho^{2}d\theta^{2}+\frac{\Sigma^{2}}{\rho^{2}}\sin^{2}\theta\,(d\phi-\omega\,dt)^{2},
\label{eq:kerr-metric}
\end{equation}
with
\begin{equation}
\omega=\frac{2aMr}{\Sigma^{2}},\quad
\rho^{2}=r^{2}+a^{2}\cos^{2}\theta,\quad
\Delta=r^{2}-2Mr+a^{2},\quad
\Sigma^{2}=(r^{2}+a^{2})^{2}-a^{2}\sin^{2}\theta\,\Delta ,
\label{eq:kerr-functions}
\end{equation}
$M$ and $a$ being the mass and the angular momentum per unit mass of the
black hole. Introducing an orthonormal vierbein $e^{a}_{\ \mu}$ via
$g_{\mu\nu}=\eta_{ab}e^{a}_{\ \mu}e^{b}_{\ \nu}$ and the antisymmetric
combination $U^{ab}_{\ \ \mu\nu}=e^{a}_{\ \mu}e^{b}_{\ \nu}-e^{a}_{\ \nu}e^{b}_{\ \mu}$,
the Weyl tensor of the Kerr metric reduces to the compact bilinear form~\cite{Daniels:1993yi,Shore:1995fz,Daniels:1995yw,Chen:2016hil}
\begin{align}
C_{\mu\nu\rho\sigma}=\,&2\mathcal{A}\Big(U^{01}_{\mu\nu}U^{01}_{\rho\sigma}-U^{23}_{\mu\nu}U^{23}_{\rho\sigma}-U^{03}_{\mu\nu}U^{03}_{\rho\sigma}+U^{12}_{\mu\nu}U^{12}_{\rho\sigma}\Big)\notag\\
&+2\mathcal{B}\Big(U^{02}_{\mu\nu}U^{02}_{\rho\sigma}-U^{13}_{\mu\nu}U^{13}_{\rho\sigma}-U^{03}_{\mu\nu}U^{03}_{\rho\sigma}+U^{12}_{\mu\nu}U^{12}_{\rho\sigma}\Big)\notag\\
&+\mathcal{C}\Big(U^{01}_{\mu\nu}U^{23}_{\rho\sigma}+U^{23}_{\mu\nu}U^{01}_{\rho\sigma}-U^{03}_{\mu\nu}U^{12}_{\rho\sigma}-U^{12}_{\mu\nu}U^{03}_{\rho\sigma}\Big)\notag\\
&+\mathcal{D}\Big({-}U^{02}_{\mu\nu}U^{13}_{\rho\sigma}-U^{13}_{\mu\nu}U^{02}_{\rho\sigma}-U^{03}_{\mu\nu}U^{12}_{\rho\sigma}-U^{12}_{\mu\nu}U^{03}_{\rho\sigma}\Big)\notag\\
&+\mathcal{E}\Big(U^{01}_{\mu\nu}U^{02}_{\rho\sigma}+U^{02}_{\mu\nu}U^{01}_{\rho\sigma}+U^{13}_{\mu\nu}U^{23}_{\rho\sigma}+U^{23}_{\mu\nu}U^{13}_{\rho\sigma}\Big)\notag\\
&+\mathcal{F}\Big(U^{01}_{\mu\nu}U^{13}_{\rho\sigma}+U^{13}_{\mu\nu}U^{01}_{\rho\sigma}-U^{02}_{\mu\nu}U^{23}_{\rho\sigma}-U^{23}_{\mu\nu}U^{02}_{\rho\sigma}\Big),
\label{eq:weyl-kerr}
\end{align}
where the six scalar functions $\mathcal{A},\ldots,\mathcal{F}$ depend
only on $r$, $\theta$, $M$ and $a$; their explicit form is given by
\begin{align}
\mathcal{A} &= \frac{Mr}{\rho^{6}\Sigma^{2}}
\left(r^{2}-3a^{2}\cos^{2}\theta\right)
\left[2\left(r^{2}+a^{2}\right)^{2}
+a^{2}\Delta\sin^{2}\theta\right],\\
\mathcal{B} &= -\frac{Mr}{\rho^{6}\Sigma^{2}}
\left(r^{2}-3a^{2}\cos^{2}\theta\right)
\left[\left(r^{2}+a^{2}\right)^{2}
+2a^{2}\Delta\sin^{2}\theta\right],\\
\mathcal{C} &= -\frac{aM\cos\theta}{\rho^{6}\Sigma^{2}}
\left(3r^{2}-a^{2}\cos^{2}\theta\right)
\left[2\left(r^{2}+a^{2}\right)^{2}
+a^{2}\Delta\sin^{2}\theta\right],\\
\mathcal{D} &= \frac{aM\cos\theta}{\rho^{6}\Sigma^{2}}
\left(3r^{2}-a^{2}\cos^{2}\theta\right)
\left[\left(r^{2}+a^{2}\right)^{2}
+2a^{2}\Delta\sin^{2}\theta\right],\\
\mathcal{E }&= -\frac{3a^{2}M\sqrt{\Delta}\cos\theta}
{\rho^{6}\Sigma^{2}}
\left(3r^{2}-a^{2}\cos^{2}\theta\right)
\left(r^{2}+a^{2}\right)\sin\theta,\\
\mathcal{F }&= \frac{3aMr\sqrt{\Delta}}
{\rho^{6}\Sigma^{2}}
\left(3r^{2}-a^{2}\cos^{2}\theta\right)
\left(r^{2}+a^{2}\right)\sin\theta.
\end{align}
Compared with the Schwarzschild case, rotation introduces the additional structures
$\mathcal{C}$-$\mathcal{F}$, which vanish when $a=0$ and which couple
the radial and polar momentum components of the photon.

\subsection{Polarization decomposition and the light-cone condition}

To organize the photon equation of motion~\eqref{eq:eom-photon} in the
Kerr background, one introduces the momentum-projected combinations \cite{Daniels:1993yi,Shore:1995fz,Daniels:1995yw}
\begin{equation}
l_{\nu}=k^{\mu}U^{01}_{\ \ \mu\nu},\qquad
n_{\nu}=k^{\mu}U^{02}_{\ \ \mu\nu},\qquad
r_{\nu}=k^{\mu}U^{03}_{\ \ \mu\nu},
\label{eq:lnr}
\end{equation}
together with the dependent combinations $p_{\nu}=k^{\mu}U^{12}_{\ \ \mu\nu}$,
$m_{\nu}=k^{\mu}U^{23}_{\ \ \mu\nu}$ and $q_{\nu}=k^{\mu}U^{13}_{\ \ \mu\nu}$,
all orthogonal to $k^{\nu}$. Contracting
Eq.~\eqref{eq:eom-photon} with $l^{\nu}$, $n^{\nu}$ and $r^{\nu}$
reduces the photon equation of motion to a homogeneous linear system for
the three independent polarization amplitudes $a\!\cdot\! l$, $a\!\cdot\! n$
and $a\!\cdot\! r$,
\begin{equation}
\begin{pmatrix} K_{11} & K_{12} & K_{13}\\ K_{21} & K_{22} & K_{23}\\ K_{31} & K_{32} & K_{33}\end{pmatrix}
\begin{pmatrix} a\cdot l\\ a\cdot n\\ a\cdot r\end{pmatrix}=0 ,
\label{eq:Kmatrix}
\end{equation}
whose coefficients $K_{ij}$ depend on $\mathcal{A},\ldots,\mathcal{F}$,
the vierbein components and $k_{\mu}$; their lengthy explicit form is
given by 
\begin{align}
K_{11}={}&
\left[-(e^{0}_{t})^{2}+(e^{3}_{t})^{2}\right]k^{t}k^{t}
+(e^{1}_{r})^{2}k^{r}k^{r}
+(e^{2}_{\theta})^{2}k^{\theta}k^{\theta}
+2e^{3}_{\phi}e^{3}_{t}k^{t}k^{\phi}
+(e^{3}_{\phi})^{2}k^{\phi}k^{\phi}
\nonumber\\
&
-16\alpha \mathcal{A}
\left[
(e^{0}_{t})^{2}k^{t}k^{t}
-(e^{1}_{r})^{2}k^{r}k^{r}
+(e^{2}_{\theta})^{2}k^{\theta}k^{\theta}
\right]
+16\alpha \mathcal{B}
\left[
(e^{3}_{t}k^{t}+e^{3}_{\phi}k^{\phi})^{2}
-(e^{2}_{\theta})^{2}k^{\theta}k^{\theta}
\right]
\nonumber\\
&
+8\alpha
e^{1}_{r}e^{2}_{\theta}k^{r}k^{\theta}
\frac{e^{3}_{t}k^{t}+e^{3}_{\phi}k^{\phi}}
     {e^{0}_{t}k^{t}}
(\mathcal{C}+2\mathcal{D})
+8\alpha \mathcal{E}
e^{1}_{r}e^{2}_{\theta}k^{r}k^{\theta}
+8\alpha
\frac{e^{3}_{t}k^{t}+e^{3}_{\phi}k^{\phi}}
     {e^{0}_{t}k^{t}}
\mathcal{F}
\left[
2(e^{0}_{t})^{2}k^{t}k^{t}
-(e^{1}_{r})^{2}k^{r}k^{r}
\right],
\\[1ex]
K_{12}={}&
16\alpha(\mathcal{A}+2\mathcal{B})e^{1}_{r}k^{r}e^{2}_{\theta}k^{\theta}-8\alpha \mathcal{E}
\left[
(e^{3}_{t}k^{t}+e^{3}_{\phi}k^{\phi})^{2}
-(e^{1}_{r})^{2}k^{r}k^{r}
+(e^{0}_{t})^{2}k^{t}k^{t}
\right]
\nonumber\\
&
+8\alpha
\frac{e^{3}_{t}k^{t}+e^{3}_{\phi}k^{\phi}}
     {e^{0}_{t}k^{t}}
\Bigl[
\mathcal{C}\left(
(e^{0}_{t})^{2}k^{t}k^{t}
-2(e^{1}_{r})^{2}k^{r}k^{r}
\right)
-\mathcal{D}\left(
(e^{0}_{t})^{2}k^{t}k^{t}
+(e^{1}_{r})^{2}k^{r}k^{r}
\right)
+\mathcal{F}e^{1}_{r}k^{r}e^{2}_{\theta}k^{\theta}
\Bigr],
\\[1ex]
K_{13}={}&
-8\alpha
(e^{3}_{t}k^{t}+e^{3}_{\phi}k^{\phi})
\left[
2(\mathcal{A}+2\mathcal{B})e^{1}_{r}k^{r}
-\mathcal{E}e^{2}_{\theta}k^{\theta}
\right]
+\frac{8\alpha}{e^{0}_{t}k^{t}}
\Bigl[
\mathcal{C}e^{2}_{\theta}k^{\theta}
\left(
-2(e^{0}_{t})^{2}k^{t}k^{t}
+(e^{1}_{r})^{2}k^{r}k^{r}
\right)
\nonumber\\
&
-\mathcal{D}e^{2}_{\theta}k^{\theta}
\left(
(e^{0}_{t})^{2}k^{t}k^{t}
+(e^{1}_{r})^{2}k^{r}k^{r}
\right)
+\mathcal{F}e^{1}_{r}k^{r}
\left[
-(e^{0}_{t})^{2}k^{t}k^{t}
+(e^{1}_{r})^{2}k^{r}k^{r}
-(e^{2}_{\theta})^{2}k^{\theta}k^{\theta}
\right]
\Bigr],\\
K_{21}={}&
16\alpha(2\mathcal{A}+\mathcal{B})e^{1}_{r}e^{2}_{\theta}k^{r}k^{\theta}-8\alpha \mathcal{E}
\left[
(e^{3}_{t}k^{t}+e^{3}_{\phi}k^{\phi})^{2}
+(e^{0}_{t})^{2}k^{t}k^{t}
-(e^{2}_{\theta})^{2}k^{\theta}k^{\theta}
\right]
\nonumber\\
&
+8\alpha
\frac{e^{3}_{t}k^{t}+e^{3}_{\phi}k^{\phi}}
     {e^{0}_{t}k^{t}}
\Bigl[
\mathcal{C}\left(
(e^{0}_{t})^{2}k^{t}k^{t}
+(e^{2}_{\theta})^{2}k^{\theta}k^{\theta}
\right)
+\mathcal{D}\left(
-(e^{0}_{t})^{2}k^{t}k^{t}
+2(e^{2}_{\theta})^{2}k^{\theta}k^{\theta}
\right)
-\mathcal{F} e^{1}_{r}e^{2}_{\theta}k^{r}k^{\theta}
\Bigr],
\\
K_{22}={}&
\left[
-(e^{0}_{t})^{2}
+(e^{3}_{t})^{2}
\right]k^{t}k^{t}
+(e^{1}_{r})^{2}k^{r}k^{r}
+(e^{2}_{\theta})^{2}k^{\theta}k^{\theta}
+2e^{3}_{\phi}e^{3}_{t}k^{t}k^{\phi}
+(e^{3}_{\phi})^{2}k^{\phi}k^{\phi}
\nonumber\\
&
+16\alpha \mathcal{A}
\left[
(e^{3}_{t}k^{t}+e^{3}_{\phi}k^{\phi})^{2}
-(e^{1}_{r})^{2}k^{r}k^{r}
\right]
-16\alpha \mathcal{B}
\left[
(e^{0}_{t})^{2}k^{t}k^{t}
+(e^{1}_{r})^{2}k^{r}k^{r}
-(e^{2}_{\theta})^{2}k^{\theta}k^{\theta}
\right]
\nonumber\\
&
-8\alpha
e^{1}_{r}e^{2}_{\theta}k^{r}k^{\theta}
\frac{e^{3}_{t}k^{t}+e^{3}_{\phi}k^{\phi}}
     {e^{0}_{t}k^{t}}
(2\mathcal{C}+\mathcal{D})
+8\alpha \mathcal{E}
e^{1}_{r}e^{2}_{\theta}k^{r}k^{\theta}
-8\alpha
\frac{e^{3}_{t}k^{t}+e^{3}_{\phi}k^{\phi}}
     {e^{0}_{t}k^{t}}
\mathcal{F}
\left[
2(e^{0}_{t})^{2}k^{t}k^{t}
-(e^{2}_{\theta})^{2}k^{\theta}k^{\theta}
\right],
\\
K_{23}={}&
-8\alpha
(e^{3}_{t}k^{t}+e^{3}_{\phi}k^{\phi})
\left[
2(2\mathcal{A}+\mathcal{B})e^{2}_{\theta}k^{\theta}
-\mathcal{E}e^{1}_{r}k^{r}
\right]
+\frac{8\alpha}{e^{0}_{t}k^{t}}
\Bigl[
\mathcal{C}e^{1}_{r}k^{r}
\left(
(e^{0}_{t})^{2}k^{t}k^{t}
+(e^{2}_{\theta})^{2}k^{\theta}k^{\theta}
\right)\nonumber\\
&+\mathcal{D}e^{1}_{r}k^{r}
\left(
2(e^{0}_{t})^{2}k^{t}k^{t}
-(e^{2}_{\theta})^{2}k^{\theta}k^{\theta}
\right)
+\mathcal{F}e^{2}_{\theta}k^{\theta}
\left[
(e^{0}_{t})^{2}k^{t}k^{t}
+(e^{1}_{r})^{2}k^{r}k^{r}
-(e^{2}_{\theta})^{2}k^{\theta}k^{\theta}
\right]
\Bigr],\\
K_{31}={}&
16\alpha
(e^{3}_{t}k^{t}+e^{3}_{\phi}k^{\phi})
\frac{e^{1}_{r}k^{r}}{e^{0}_{t}k^{t}}
\Bigl[
\mathcal{A}(e^{0}_{t}k^{t}-e^{2}_{\theta}k^{\theta})
-\mathcal{B}(e^{0}_{t}k^{t}+e^{2}_{\theta}k^{\theta})
+\mathcal{E}e^{2}_{\theta}k^{\theta}
\Bigr]
\nonumber\\
&
+8\alpha
\frac{e^{2}_{\theta}k^{\theta}}{e^{0}_{t}k^{t}}
\Bigl[
\mathcal{C}\!\left(
-2(e^{0}_{t})^{2}k^{t}k^{t}
+(e^{3}_{t}k^{t}+e^{3}_{\phi}k^{\phi})^{2}
\right)
+\mathcal{D}\!\left(
-(e^{0}_{t})^{2}k^{t}k^{t}
+2(e^{3}_{t}k^{t}+e^{3}_{\phi}k^{\phi})^{2}
\right)
\Bigr]
\nonumber\\
&
-8\alpha
\frac{e^{1}_{r}k^{r}}{e^{0}_{t}k^{t}}
\mathcal{F}
\left[
(e^{0}_{t})^{2}k^{t}k^{t}
+(e^{3}_{t}k^{t}+e^{3}_{\phi}k^{\phi})^{2}
\right],
\\
K_{32}={}&
16\alpha
(e^{3}_{t}k^{t}+e^{3}_{\phi}k^{\phi})
\frac{1}{e^{0}_{t}k^{t}}
\Bigl[
\mathcal{A}\!\left(
(e^{1}_{r})^{2}k^{r}k^{r}
-e^{0}_{t}k^{t}e^{2}_{\theta}k^{\theta}
\right)
+\mathcal{B}\!\left(
(e^{1}_{r})^{2}k^{r}k^{r}
+e^{0}_{t}k^{t}e^{2}_{\theta}k^{\theta}
\right)
+\mathcal{E}e^{1}_{r}k^{r}
\Bigr]
\nonumber\\
&
+8\alpha
\frac{e^{1}_{r}k^{r}}{e^{0}_{t}k^{t}}
\Bigl[
\mathcal{C}\!\left(
(e^{0}_{t})^{2}k^{t}k^{t}
-2(e^{3}_{t}k^{t}+e^{3}_{\phi}k^{\phi})^{2}
\right)
+\mathcal{D}\!\left(
2(e^{0}_{t})^{2}k^{t}k^{t}
-(e^{3}_{t}k^{t}+e^{3}_{\phi}k^{\phi})^{2}
\right)
\Bigr]
\nonumber\\
&
+8\alpha
\frac{e^{2}_{\theta}k^{\theta}}{e^{0}_{t}k^{t}}
\mathcal{F}
\left[
(e^{0}_{t})^{2}k^{t}k^{t}
+(e^{3}_{t}k^{t}+e^{3}_{\phi}k^{\phi})^{2}
\right],\\
K_{33}={}&
\left[
-(e^{0}_{t})^{2}
+(e^{3}_{t})^{2}
\right]k^{t}k^{t}
+(e^{1}_{r})^{2}k^{r}k^{r}
+(e^{2}_{\theta})^{2}k^{\theta}k^{\theta}
+2e^{3}_{\phi}e^{3}_{t}k^{t}k^{\phi}
+(e^{3}_{\phi})^{2}k^{\phi}k^{\phi}
\nonumber\\
&
+16\alpha \mathcal{A}
\left[
(e^{0}_{t})^{2}k^{t}k^{t}
+(e^{2}_{\theta})^{2}k^{\theta}k^{\theta}
-(e^{3}_{t}k^{t}+e^{3}_{\phi}k^{\phi})^{2}
\right]
+16\alpha \mathcal{B}
\left[
(e^{0}_{t})^{2}k^{t}k^{t}
+(e^{1}_{r})^{2}k^{r}k^{r}
-(e^{3}_{t}k^{t}+e^{3}_{\phi}k^{\phi})^{2}
\right]
\nonumber\\
&
-8\alpha
\frac{e^{3}_{t}k^{t}+e^{3}_{\phi}k^{\phi}}
     {e^{0}_{t}k^{t}}
\Bigl[
e^{1}_{r}k^{r}
\Bigl(
\mathcal{C}(e^{0}_{t}k^{t}-e^{2}_{\theta}k^{\theta})
+\mathcal{D}(e^{0}_{t}k^{t}+e^{2}_{\theta}k^{\theta})
\Bigr)
-\mathcal{F}\Bigl(
(e^{1}_{r})^{2}k^{r}k^{r}
-(e^{2}_{\theta})^{2}k^{\theta}k^{\theta}
\Bigr)
\Bigr]
\nonumber\\
&
-16\alpha \mathcal{E}\,e^{1}_{r}k^{r}e^{2}_{\theta}k^{\theta}.
\end{align}
in Ref.~\cite{Chen:2016hil}. A non-trivial polarization requires $\det K=0$, which in general must be solved numerically because of the complexity introduced by rotation.

A simplification occurs for a photon confined to the equatorial plane
($k^{\theta}=0$, $\theta=\pi/2$), for which the off-diagonal terms
$K_{12}$, $K_{21}$, $K_{23}$ and the functions $\mathcal{C}$,
$\mathcal{D}$, $\mathcal{E}$ vanish identically, and the determinant
condition factorizes,
\begin{equation}
\det K=\tilde K_{11}\,\tilde K_{22}\,\tilde K_{33}=0 .
\label{eq:Kdet-eq}
\end{equation}
The first root, $\tilde K_{11}=0$, simply reproduces the uncoupled light
cone $k^{\mu}k_{\mu}=0$ and corresponds to an unphysical polarization
that is discarded, as in the non-rotating
case~\cite{Drummond:1979pp,Chen:2015cpa,Lu:2016gsf}. The second root, $\tilde
K_{22}=0$, namely
\begin{equation}
-(1+16\alpha\mathcal{B})(e^{0}_{\ t})^{2}k^{t}k^{t}
+\big[1-16\alpha(\mathcal{A}+\mathcal{B})\big](e^{1}_{\ r})^{2}k^{r}k^{r}
+(1+16\alpha\mathcal{A})(e^{3}_{\ t}k^{t}+e^{3}_{\ \phi}k^{\phi})^{2}
-16\alpha\mathcal{F}e^{0}_{\ t}k^{t}(e^{3}_{\ t}k^{t}+e^{3}_{\ \phi}k^{\phi})=0,
\label{eq:PPM}
\end{equation}
governs a photon whose polarization vector is (effectively) aligned with
$m_{\mu}$, i.e.\ orthogonal to the orbital plane; following
Ref.~\cite{Chen:2016hil} we denote this mode \emph{PPM}.
The third root, $\tilde K_{33}=0$,
\begin{align}
&\big[1-16\alpha(\mathcal{A}+\mathcal{B})\big]\Big[-(1+16\alpha\mathcal{A})(e^{0}_{\ t})^{2}k^{t}k^{t}+(1+16\alpha\mathcal{B})(e^{3}_{\ t}k^{t}+e^{3}_{\ \phi}k^{\phi})^{2}+16\alpha\mathcal{F}e^{0}_{\ t}k^{t}(e^{3}_{\ t}k^{t}+e^{3}_{\ \phi}k^{\phi})\Big]\notag\\
&\quad+\Big[1+16\alpha(\mathcal{A}+\mathcal{B})+256\alpha^{2}\mathcal{A}\mathcal{B}+64\alpha^{2}\mathcal{F}^{2}\Big](e^{1}_{\ r})^{2}k^{r}k^{r}=0,
\label{eq:PPL}
\end{align}
corresponds to a polarization vector lying in the plane of motion, which
we denote \emph{PPL}, following the notation of
Refs.~\cite{Chen:2015cpa,Lu:2016gsf}.

Equations~\eqref{eq:PPM} and \eqref{eq:PPL} are the two physical
light-cone conditions for a Weyl-coupled photon moving in the equatorial
plane of a Kerr black hole: they define two distinct effective metrics
for the two polarization states, so that PPL and PPM photons emitted
from the same source generically follow different trajectories --- a
gravitational analogue of birefringence already present in the
Schwarzschild case~\cite{Chen:2015cpa} and now shown to persist, in a
more intricate form, once rotation is included. Both conditions reduce
to the ordinary Kerr light cone, with no distinction between
polarizations, in the limit $\alpha\to0$. These two light-cone
conditions are the starting point for the construction of the effective
optical metric and the analysis of strong gravitational lensing
presented in Ref.~\cite{Chen:2016hil}.

Let us now study the strong gravitational lensing by a Kerr black hole as photon couples to Weyl tensor.
Actually, the light cone conditions \eqref{eq:PPM} and \eqref{eq:PPL} indicate that the motion of the coupled photons in the
equatorial plane is non-geodesic in the Kerr metric. However, these photons in the equatorial plane can be
looked as moving along the null geodesics of the effective metric $\gamma_{\mu\nu}$, i.e.,
\begin{equation}
\gamma_{\mu\nu}k^\mu k^\nu=0.
\end{equation}
The effective metric for the coupled photon limited on the equatorial plane in a Kerr black hole spacetime
can be expressed as
\begin{equation}
ds^2=-A(r)dt^2+B(r)dr^2+C(r)d\phi^2-2D(r)dtd\phi,
\end{equation}
where $D(r) = -g_{t\phi}$ encodes the frame-dragging effect.

The metric coefficients $A(r)$, $B(r)$, $C(r)$ and $D(r)$ depend on the polarization directions of the
coupled photon. For the PPL, the functions $A(r)$, $B(r)$, $C(r)$ and $D(r)$ can be expressed as
\begin{align}
A(r)
&=
\frac{
r^{10}(r-2M)\left[(r^{3}+2Ma^{2}+a^{2}r)^{2}-16\alpha Mr^{3}\right]+W_{3}
}{
r^{8}(r^{3}+2Ma^{2}+a^{2}r)^{2}(r^3+8\alpha M)+W_4+8\alpha MW_1},
\nonumber\\
B(r)
&=
\frac{
r^{10}(r^{3}+2Ma^{2}+a^{2}r)^{2}}{(r^{2}-2Mr+a^{2})
\left[
r^{5}(r^{3}+2Ma^{2}+a^{2}r)^{2}(r^{3}+8\alpha M)+W_{1}
\right]
},
\nonumber\\
C(r)
&=
\frac{
r^{6}(r^{3}+2Ma^{2}+a^{2}r)^{2}
\left[
r^{4}(r^{3}+2Ma^{2}+a^{2}r)
+8\alpha M(2r^{4}+5a^{2}r^{2}-2a^{2}Mr+3a^{4})
\right]
}{
(r^{3}-8\alpha M)
\left[
r^{5}(r^{3}+2Ma^{2}+a^{2}r)^{2}(r^{3}+8\alpha M)+W_{1}
\right]
},
\nonumber\\
D(r)
&=
\frac{
2aMr^{6}(r^{3}+2Ma^{2}+a^{2}r)
\left[
r^{4}(r^{3}+2Ma^{2}+a^{2}r)+W_{5}
\right]
}{
(r^{3}-8\alpha M)
\left[
r^{5}(r^{3}+2Ma^{2}+a^{2}r)^{2}(r^{3}+8\alpha M)+W_{1}
\right]
}.
\end{align}
with
\begin{align}
W_{3}
&=
-8\alpha Ma^{2}r^{6}
\Bigl[
r^{4}(5r^{2}+6Mr-20M^{2})
+a^{2}r(7r^{3}+26Mr^{2}-16M^{2}r-8M^{3})
+3a^{4}(r^{2}+6Mr+4M^{2})
\Bigr],
\nonumber\\
W_{4}
&=
-48\alpha^{2}M^{2}r^{3}
\Bigl[
4r^{8}
-a^{2}r^{5}(7r-46M)
-a^{4}r^{2}(41r^{2}-76Mr-16M^{2})
-15a^{6}r(3r-2M)
-15a^{8}
\Bigr],
\nonumber\\
W_{5}
&=
2\alpha
\Bigl[
r^{4}(9r-10M)
+2a^{2}r(9r^{2}+Mr-4M^{2})
+3a^{4}(3r+4M)
\Bigr],
\end{align}

While for PPM, these functions become
\begin{align}
A(r)
&=
\frac{
r^{7}(r-2M)\left[(r^{3}+2Ma^{2}+a^{2}r)^{2}+16\alpha Mr^{3}\right]+W_{0}
}{
r^{5}(r^{3}+2Ma^{2}+a^{2}r)^{2}(r^{3}+8\alpha M)+W_{1}
},
\nonumber\\
B(r)
&=
\frac{r^{5}}{
r^{5}(r^{2}-2Mr+a^{2})(r^{3}-8\alpha M)
},
\nonumber\\
C(r)
&=
\frac{
r^{3}(r^{3}+2Ma^{2}+a^{2}r)^{2}
\left[
r^{4}(r^{3}+2Ma^{2}+a^{2}r)
-8\alpha M(r^{4}+4a^{2}r^{2}-4a^{2}Mr+3a^{4})
\right]
}{
r^{5}(r^{3}+2Ma^{2}+a^{2}r)^{2}(r^{3}+8\alpha M)+W_{1}
},
\nonumber\\
D(r)
&=
\frac{
2aMr^{3}(r^{3}+2Ma^{2}+a^{2}r)
\left[
r^{4}(r^{3}+2Ma^{2}+a^{2}r)+W_{2}
\right]
}{
r^{5}(r^{3}+2Ma^{2}+a^{2}r)^{2}(r^{3}+8\alpha M)+W_{1}
},
\end{align}
with
\begin{align}
W_{0}
&=
8\alpha Ma^{2}r^{3}
\Bigl[
r^{4}(7r^{2}+6Mr-28M^{2})
+4a^{2}r(2r^{3}+7Mr^{2}-5M^{2}r-4M^{3})
+3a^{4}(r^{2}+6Mr+4M^{2})
\Bigr],
\nonumber\\
W_{1}
&=
-16\alpha^{2}M^{2}
\Bigl[
8r^{8}
-a^{2}r^{5}(29r-122M)
-a^{4}r^{2}(127r^{2}-212Mr-32M^{2})
-45a^{6}r(3r-2M)
-45a^{8}
\Bigr],
\nonumber\\
W_{2}
&=
-2\alpha
\Bigl[
r^{4}(9r-14M)
-2a^{2}r(9r^{2}-Mr-8M^{2})
+3a^{4}(3r+4M)
\Bigr].
\end{align}

In this work, we focus on photons propagating in the retrograde direction with respect to the rotation of the black hole because they possess a larger photon orbit radius than prograde ones. As a result, the corresponding strong-lensing observables are less affected by the near-horizon region and provide a complementary regime for probing the EFT corrections. For a Kerr black hole with positive spin parameter $a>0$, retrograde photon trajectories are characterized by a negative conserved angular momentum J. By choosing the affine parameter such that the conserved quantity associated with time-translation symmetry is normalized to $E=-1$, the impact parameter is given by
\begin{align}
u\equiv \frac{J}{E}=-J<0,
\end{align}
where $E$ and $J$ are the conserved energy and axial angular momentum of the photon, respectively. Since the strong gravitational lensing observables are determined by the properties of the unstable circular photon orbit, we first evaluate the corresponding photon sphere radius. In the Kerr background, the radius of the retrograde photon orbit is given by
\begin{align}
r_{\mathrm{ph}}^{(0)}
=
2M\left[
1+\cos\!\left(
\frac{2}{3}\arccos\frac{a}{M}
\right)
\right].
\end{align}
Treating the higher-curvature corrections perturbatively, the photon sphere radius can be expanded as
\begin{align}
r_{\mathrm{ph}}
=
r_{\mathrm{ph}}^{(0)}
+\delta r_{\mathrm{ph}}
+\mathcal{O}(\alpha^2),
\end{align}
where $\delta r_{\mathrm{ph}}$ denotes the leading-order EFT correction. Solving the circular null orbit condition to first order in the EFT couplings, we obtain
\begin{align}
r_{\rm ph}=&2M(1+c)
-\alpha\Bigl(
\dfrac{6M^2-20M^2(1+c)+12M^2(1+c)^2
-\dfrac{a M(-2M+4M(1+c))}{S}}{M(1+c)(-2M+2M(1+c))}\notag\\
&-\dfrac{-2a^2 M+12M^3(1+c)-20M^3(1+c)^2+8M^3(1+c)^3-2a M S}{M(1+c)(-2M+2M(1+c))^2}\notag\\
&-\dfrac{-2a^2 M+12M^3(1+c)-20M^3(1+c)^2+8M^3(1+c)^3-2a M S}{2M^2(1+c)^2(-2M+2M(1+c))}\Bigr)^{-1}
\notag \\
&\times\Biggl[
\frac{1}{16M^6(1+c)^7\mathcal{D}^2}\Bigl(
-144a^8 M^3
-504a^8 M^3(1+c)
+288a^6 M^5(1+c) \notag \\
&-528a^8 M^3(1+c)^2
+768a^6 M^5(1+c)^2
-120a^8 M^3(1+c)^3
-1632a^6 M^5(1+c)^3 \notag \\
&-5280a^6 M^5(1+c)^4
+3456a^4 M^7(1+c)^4
-1632a^6 M^5(1+c)^5
+6912a^4 M^7(1+c)^5 \notag \\
&-16128a^4 M^7(1+c)^6
-8064a^4 M^7(1+c)^7
+20480a^2 M^9(1+c)^7
-9216a^2 M^9(1+c)^8 \notag \\
&-16384a^2 M^9(1+c)^9
+20480M^{11}(1+c)^{10}
-18432M^{11}(1+c)^{11}
\Bigr) \notag \\
&-\frac{1}{16M^6(1+c)^7\mathcal{D}^2}\Bigl(
-144a^8 M^3
-192a^8 M^3(1+c)
+192a^6 M^5(1+c)
+96a^8 M^3(1+c)^2 \notag \\
&-192a^6 M^5(1+c)^2
+120a^8 M^3(1+c)^3
-1824a^6 M^5(1+c)^3
+672a^6 M^5(1+c)^4 \notag \\
&+1152a^4 M^7(1+c)^4
+1632a^6 M^5(1+c)^5
-5760a^4 M^7(1+c)^5
+8064a^4 M^7(1+c)^7 \notag \\
&-5632a^2 M^9(1+c)^7
-8192a^2 M^9(1+c)^8
+16896a^2 M^9(1+c)^9 \notag \\
&-16384M^{11}(1+c)^{10}
+16384M^{11}(1+c)^{11}
\Bigr) \notag \\
&+512M^8(1+c)^8\mathcal{D}^2\Biggl(
\frac{a}{256M^8(1+c)^9(-2M+2M(1+c))\mathcal{K}^2} \notag \\
&\times\Bigl(
-\frac{1}{16M^5(1+c)^6 S\,\mathcal{D}}\Bigl(24a^4 M^2-24a^6 M^2+36a^4 M^2(1+c)-36a^6 M^2(1+c)+16a^4 M^4(1+c)\notag\\
&+16a^4 M^4(1+c)^2+64a^2 M^6(1+c)^4+256M^8(1+c)^7-256M^8(1+c)^8\Bigr)\notag \\
&-\frac{a\bigl(3a^2(4M+6M(1+c))+16M^4(1+c)^4(-10M+18M(1+c))+4a^2 M(1+c)(-4M^2+2M^2(1+c)+36M^2(1+c)^2)\bigr)}{8M^4(1+c)^5\mathcal{D}}
\Bigr) \notag \\
&+\Bigl(-\frac{a}{1+c}-S\Bigr)
\Biggl(
\frac{a}{4096M^{13}(1+c)^{14}(-2M+2M(1+c))\mathcal{D}^3\mathcal{K}^2} \notag \\
&\times\Bigl(
-384a^6 M^4+384a^8 M^4
-816a^6 M^4(1+c)+1440a^8 M^4(1+c)-384a^6 M^6(1+c) \notag \\
&+144a^6 M^4(1+c)^2+1632a^8 M^4(1+c)^2-2304a^6 M^6(1+c)^2 \notag \\
&+1296a^6 M^4(1+c)^3+384a^8 M^4(1+c)^3-4224a^4 M^6(1+c)^3+1152a^6 M^6(1+c)^3 \notag \\
&+720a^6 M^4(1+c)^4-4800a^4 M^6(1+c)^4+14208a^6 M^6(1+c)^4-4608a^4 M^8(1+c)^4 \notag \\
&+5952a^4 M^6(1+c)^5+8832a^6 M^6(1+c)^5-29184a^4 M^8(1+c)^5 \notag \\
&+6912a^4 M^6(1+c)^6+3456a^6 M^6(1+c)^6-10752a^2 M^8(1+c)^6+15360a^4 M^8(1+c)^6 \notag \\
&-1536a^2 M^8(1+c)^7+31488a^4 M^8(1+c)^7-39936a^2 M^{10}(1+c)^7 \notag \\
&+16128a^2 M^8(1+c)^8+39168a^4 M^8(1+c)^8-21504a^2 M^{10}(1+c)^8 \notag \\
&-44032a^2 M^{10}(1+c)^9+129024a^2 M^{10}(1+c)^{10} \notag \\
&-12288M^{12}(1+c)^{10}-106496M^{12}(1+c)^{11}+110592M^{12}(1+c)^{12}
\Bigr) \notag \\
&+\frac{a}{2M(1+c)^2}\Bigl(
\frac{1}{128M^9(1+c)^{10}(-2M+2M(1+c))\mathcal{K}^2}-\frac{1}{131072M^{17}(1+c)^{17}(-2M+2M(1+c))^2\mathcal{K}^4} \notag \\
&\Bigl(-131072M^{14}(1+c)^{13}(-2M+2M(1+c))-512a^2 M^7(1+c)^6\bigl(3a^4(4M^2+12M^2(1+c
+4M^2(1+c)^2)\notag\\
&+16M^4(1+c)^4(-20M^2+12M^2(1+c)+20M^2(1+c)^2)\notag\\
&+2a^2 M(1+c)(-8M^3-32M^3(1+c)+104M^3(1+c)^2+56M^3(1+c)^3)\bigr)\Bigr)
\Bigr)
\Biggr)
\Biggr)
\Biggr]\notag\\
&\quad\text{for PPL in Kerr spacetime}, \\
r_{\rm ph}=&2M(1+c)
-\alpha
\Bigl(\dfrac{6M^2-20M^2(1+c)+12M^2(1+c)^2-\dfrac{aM(-2M+4M(1+c))}{S}}{M(1+c)(-2M+2M(1+c))}\notag\\
&-\dfrac{-2a^2M+12M^3(1+c)-20M^3(1+c)^2+8M^3(1+c)^3-2aMS}{M(1+c)(-2M+2M(1+c))^2}\notag\\
&-\dfrac{-2a^2M+12M^3(1+c)-20M^3(1+c)^2+8M^3(1+c)^3-2aMS}{2M^2(1+c)^2(-2M+2M(1+c))}\Bigr)^{-1}
\notag \\
&\times\Biggl[
-\frac{1}{16M^6(1+c)^7\mathcal{D}^2}\Bigl(
-144a^8M^3
-504a^8M^3(1+c)
+160a^6M^5(1+c)
-528a^8M^3(1+c)^2
+480a^6M^5(1+c)^2 \notag \\
&-120a^8M^3(1+c)^3
-1728a^6M^5(1+c)^3
-5120a^6M^5(1+c)^4
+1920a^4M^7(1+c)^4
-1536a^6M^5(1+c)^5 \notag \\
&+4992a^4M^7(1+c)^5
-15360a^4M^7(1+c)^6
-6912a^4M^7(1+c)^7
+15360a^2M^9(1+c)^7
-12288a^2M^9(1+c)^8 \notag \\
&-12288a^2M^9(1+c)^9
+4096M^{11}(1+c)^{10}
-6144M^{11}(1+c)^{11}
\Bigr) \notag \\
&+\frac{1}{16M^6(1+c)^7\mathcal{D}^2}\Bigl(
-144a^8M^3
-192a^8M^3(1+c)
+160a^6M^5(1+c)
+96a^8M^3(1+c)^2
-224a^6M^5(1+c)^2 \notag \\
&+120a^8M^3(1+c)^3
-1792a^6M^5(1+c)^3
+704a^6M^5(1+c)^4
+1152a^4M^7(1+c)^4
+1632a^6M^5(1+c)^5 \notag \\
&-5504a^4M^7(1+c)^5
+7808a^4M^7(1+c)^7
-4608a^2M^9(1+c)^7
-7168a^2M^9(1+c)^8
+14848a^2M^9(1+c)^9 \notag \\
&-8192M^{11}(1+c)^{10}
+8192M^{11}(1+c)^{11}
\Bigr) \notag \\
&+64M^5(1+c)^5\mathcal{D}^2\Biggl(
\frac{a}{32M^5(1+c)^6(-2M+2M(1+c))\mathcal{K}^2} \notag \\
&\times\frac{1}{8M^4(1+c)^5S\,\mathcal{D}}\Bigl(aS\bigl(12a^4M+18a^4M(1+c)+48a^2M^3(1+c)+24a^2M^3(1+c)^2\notag\\
&-144a^2M^3(1+c)^3-160M^5(1+c)^4+288M^5(1+c)^5\bigr)\notag \\
&+64a^4M^3+16a^4M^3(1+c)-280a^4M^3(1+c)^2+8a^4M^3(1+c)^3-32a^2M^5(1+c)^3+64a^2M^5(1+c)^5\notag\\
&-128M^7(1+c)^6+128M^7(1+c)^7\Bigr) \notag \\
&+\Bigl(-\frac{a}{1+c}-S\Bigr)\Biggl(
-\frac{a}{256M^9(1+c)^{10}(-2M+2M(1+c))\mathcal{D}^3\mathcal{K}^2} \notag \\
&\times\Bigl(
312a^8M^3-768a^6M^5
+888a^8M^3(1+c)-1600a^6M^5(1+c)
+840a^8M^3(1+c)^2+960a^6M^5(1+c)^2 \notag \\
&+360a^8M^3(1+c)^3+7200a^6M^5(1+c)^3-9216a^4M^7(1+c)^3
+4064a^6M^5(1+c)^4-12288a^4M^7(1+c)^4 \notag \\
&+1728a^6M^5(1+c)^5+37120a^4M^7(1+c)^5
+11392a^4M^7(1+c)^6-38400a^2M^9(1+c)^6 \notag \\
&-9216a^4M^7(1+c)^7+18944a^2M^9(1+c)^7
+53760a^2M^9(1+c)^8-27648a^2M^9(1+c)^9 \notag \\
&+18432M^{11}(1+c)^9-69632M^{11}(1+c)^{10}+55296M^{11}(1+c)^{11}
\Bigr) \notag \\
&+\frac{a}{2M(1+c)^2}\Bigl(
\frac{1}{16M^6(1+c)^7(-2M+2M(1+c))\mathcal{K}^2}-\frac{16384M^{11}(1+c)^{10}(-2M+2M(1+c))}{2048M^{11}(1+c)^{11}(-2M+2M(1+c))^2\mathcal{K}^4} \notag \\
&-\frac{64a^2M^4(1+c)^3\bigl(3a^4(4M^2+12M^2(1+c)+4M^2(1+c)^2)}{2048M^{11}(1+c)^{11}(-2M+2M(1+c))^2\mathcal{K}^4} \notag \\
&-\frac{+16M^4(1+c)^4(-28M^2+12M^2(1+c)+28M^2(1+c)^2)}{2048M^{11}(1+c)^{11}(-2M+2M(1+c))^2\mathcal{K}^4} \notag \\
&-\frac{+8a^2M(1+c)(-4M^3-10M^3(1+c)+28M^3(1+c)^2+16M^3(1+c)^3)\bigr)}{2048M^{11}(1+c)^{11}(-2M+2M(1+c))^2\mathcal{K}^4}
\Bigr)
\Biggr)
\Biggr)
\Biggr] \notag\\
&\quad\text{for PPM in Kerr spacetime}, 
\end{align}
where
\begin{align}
c&=\cos\left(\frac{2}{3}\arccos\frac{a}{M}\right),\\
S&=\sqrt{a^2-4M^2(1+c)+4M^2(1+c)^2},\\
\mathcal{D}&=2a^2M+2a^2M(1+c)+8M^3(1+c)^3,\\
\mathcal{K}&=8M^3(1+c)^3+2a^2M(2+c).
\end{align}

\section{Strong-Field Gravitational Lensing in Effective Field Theory}
\label{sec:strong lens}

In the previous section, we analyzed gravitational lensing in the weak-field regime, where the deflection angle can be treated perturbatively. In this section, we turn to the strong-field regime, where light rays propagate in the vicinity of the photon sphere and nonlinear effects become significant, leading to large deflection angles and the formation of relativistic images.

We consider a general stationary and axisymmetric spacetime with a non-vanishing off-diagonal component $g_{t\phi}$. The metric can be written as
\begin{equation}
ds^2 = -A(r) dt^2 + B(r) dr^2 + C(r) d\phi^2 - 2 D(r) dt d\phi.
\end{equation}

\subsection{Constants of Motion and Photon Trajectories}

Due to stationarity and axisymmetry, the energy $E$ and angular momentum $L$ are conserved:
\begin{align}
\label{time conserved quantity}
E &=  A(r)\dot{t} + D(r)\dot{\phi}, \\
-J &=  C(r)\dot{\phi}-D(r)\dot{t}.
\end{align}
Here and in what follows, an overdot denotes differentiation with respect to the affine parameter along the geodesic.

Solving for $\dot{t}$ and $\dot{\phi}$, we obtain
\begin{align}
\dot{t} &= \frac{E C+J D}{A C + D^2}, \\
\dot{\phi} &= \frac{E D -J A }{A C + D^2}.
\end{align}

The null condition $ds^2 = 0$ yields the radial equation:
\begin{align}
B\dot{r}^2 = 
\frac{E^2C +2 E J D- J^2A}{A C + D^2}.
\end{align}

By choosing a parametrization of the null geodesic in which the conserved quantity associated with time-translation symmetry is set to $-1$,we rewrite
\begin{align}
\label{dot r}
\dot{r}^2 = \frac{1}{B} 
\frac{C -2 JD - J^2A }{A C + D^2}.
\end{align}

\subsection{Divergent term of the deflection angle}

In this subsection, we review the strong deflection limit analysis of gravitational lensing developed by Valerio Bozza \cite{Bozza:2002zj}. We define new variable
\begin{align}
z = \frac{A(r) - A_0}{1 - A_0},
\end{align}
where $A_0 = A(r_0)$. The orbit equation becomes
\begin{align}
\frac{d\phi}{dr}\equiv I(r_0) = \int_{0}^{1} R(z,r_0)\, f(z,r_0)\, dz,
\end{align}
with
\begin{align}
R(z,r_0) &= \frac{2 \sqrt{BA_0}}{A'} (1 - A_0)\frac{D+JA}{\sqrt{C}\sqrt{AC+D^2}}, \\
f(z,r_0) &= \frac{1}{\sqrt{A_0-A\frac{C_0}{C} +\frac{2J}{C}(AD_0-A_0D)}}.
\end{align}
Here, all functions with the subscript $0$ are evaluated at $r_0$ and ${}^\prime$ denotes differentiation with respect to $r$.

The function $R(z, r_0)$ is regular for all values of $z$ and $r_0$, whereas $f(z, r_0)$ diverges in the limit $z \to 0$. To determine the leading behavior of this divergence in the integrand, we expand the argument of the square root in $f(z, r_0)$ up to second order in $z$:
\begin{align}
\label{pole expansion}
f(z,r_0) \sim f_0(z,r_0) = \frac{1}{\sqrt{p(r_0) z + q(r_0) z^2}}.
\end{align}

When $p(r_0)$ remains nonvanishing, the leading divergence of $f_0$ scales as $z^{-1/2}$, which is integrable and therefore leads to a finite contribution. In contrast, if $p(r_0)$ vanishes, the divergence instead scales as $z^{-1}$, causing the integral to diverge.

From the structure of $p(r_0)$, one finds that it becomes zero at $r_0 = r_{\mathrm{ph}}$, where $r_{\mathrm{ph}}$ is specified by the condition introduced in the previous section:
\begin{align}
\label{ph radius}
\frac{d}{dr}\left(\dfrac{-D(r)-\sqrt{D^2(r)+A(r)C(r)}}{A(r)}\right)\bigg|_{r=r_{\rm ph}} = 0.
\end{align}
In this work, we restrict our attention to retrograde photon trajectories, corresponding to $J <0$. At this location, the impact parameter reaches its critical value $u_c$, and photons follow unstable circular orbits around the black hole. For trajectories with $r_0 < r_{\mathrm{ph}}$, photons are captured by the central object and do not escape to infinity.

To evaluate the integral, we decompose it into two contributions:
\begin{align}
I(r_0) = I_D(r_0) + I_R(r_0),
\end{align}
where
\begin{align}
I_D(r_0) &= \int_{0}^{1} R(0,r_{ph})\, f_0(z,r_0)\, dz,
\end{align}
captures the divergent behavior, and
\begin{align}
I_R(r_0) &= \int_{0}^{1} g(z,r_0)\, dz,
\end{align}
with
\begin{align}
g(z,r_0) = R(z,r_0) f(z,r_0) - R(0,r_{\rm ph}) f_0(z,r_0),
\end{align}
represents the original integrand with the singular contribution removed.

We proceed by computing each part independently and then combining the results to reconstruct the deflection angle. In this subsection, we focus on $I_D$ and its divergence, while in the following subsection we now examine how the finite contribution is obtained.

The integral $I_D(r_0)$ admits an exact evaluation:
\begin{align}
I_D(r_0) = \frac{R(0,r_{\rm ph})}{2\sqrt{q(r_0)}} 
\log \left(\frac{\sqrt{q(r_0)} + \sqrt{p(r_0) + q(r_0)}}{\sqrt{p(r_0)}}\right).
\end{align}

Here, all functions with the subscript ${}_{\rm ph}$ are evaluated at $r_{\rm ph}$.

Substituting these expressions into $I_D(r_0)$ and reorganizing terms, we obtain
\begin{align}
I_D(r_0)= - a \log\left( \frac{r_0}{r_{\rm ph}} - 1 \right)+ b_D+ \mathcal{O}((r_0 - r_{\rm ph})\log(r_0-r_{\rm ph})),
\end{align}
with
\begin{align}
a &= \frac{R(0,r_{\rm ph})}{\sqrt{q(r_{\rm ph})}}, \\
b_D &= \frac{R(0,r_{\rm ph})}{\sqrt{q(r_{\rm ph})}}
\log \left( \frac{2(1 - A_{\rm ph})}{A'_{\rm ph} r_{\rm ph}} \right).
\end{align}

This result provides the leading divergent contribution to the deflection angle, which is logarithmic in nature, as expected.

\subsection{Regular contribution to the deflection angle}

To determine the full coefficient $b$ relevant for strong-field gravitational lensing, following the method developed by Bozza \cite{Bozza:2002zj}, we must supplement $b_D$ with the corresponding contribution from the regular part of the integral defined in Eq.~(\ref{deflection angle}).

We expand $I_R(r_0)$ in powers of $(r_0 - r_{ph})$:
\begin{align}
I_R(r_0)
=
\sum_{n=0}^{\infty}
\frac{1}{n!} (r_0 - r_{\rm ph})^n
\int_{0}^{1}
\left.
\frac{\partial^n g}{\partial r_0^n}
\right|_{r_0 = r_{\rm ph}}
dz.
\end{align}

Without subtracting the divergent part from $R(z,r_0)f(z,r_0)$, the coefficient corresponding to $n=0$ would diverge, whereas all higher-order terms would be finite. By construction, however, the function $g(z,r_0)$ is regular at $z=0$ and $r_0 = r_{ph}$, as can be verified through a series expansion, using $p(r_{\rm ph}) = 0$.

Restricting ourselves to terms up to $\mathcal{O}(r_0 - r_{\rm ph})$, we keep only the leading term:
\begin{align}
I_R(r_0)
=
\int_{0}^{1} g(z,r_{\rm ph})\, dz
+ \mathcal{O}(r_0 - r_{\rm ph}),
\end{align}
and define
\begin{align}
b_R = I_R(r_{\rm ph}).
\end{align}

Including also the $-\pi$ contribution in the deflection angle, we arrive at
\begin{align}
b = -\pi + b_D + b_R.
\end{align}

The quantity $b_R$ can be evaluated numerically for general metrics, since the integrand is regular. In several cases, however, an analytic expression can be obtained. For instance, in the Schwarzschild spacetime, the integral can be computed exactly. More broadly, one may expand the integral around the Schwarzschild limit in terms of metric parameters, allowing each term in the expansion to be determined separately.

In the general formula for the deflection angle, the impact parameter $u$ is introduced. We begin by defining it as a function of the closest approach distance $r_0$:
\begin{align}
\label{impact para}
u = \dfrac{-D_0-\sqrt{D_0^2+A_0C_0}}{A_{0}}.
\end{align}

From Eq.~(\ref{ph radius}), the minimum impact parameter is given by
\begin{align}
u_{\rm ph} = u = \dfrac{-D_{\rm ph}-\sqrt{D_{\rm ph}^2+A_{\rm ph}C_{\rm ph}}}{A_{\rm ph}}.
\end{align}

Expanding Eq.~(\ref{impact para}) yields
\begin{align}
u - u_{ph}= c (r_0 - r_{\rm ph})^2+\mathcal{O}\left( (r_0 - r_{\rm ph})^3\right).
\end{align}

Using this relation, the deflection angle may be expressed as a function of the angular position $\theta$:
\begin{align}
\alpha(\theta)
=
- a \log\left( \frac{\theta D_{OL}}{u_{\rm ph}} - 1 \right)
+ \bar{b},
\end{align}
with
\begin{align}
\bar{a} &= \frac{a}{2} = \frac{R(0,r_{\rm ph})}{2\sqrt{q(r_{\rm ph})}}, \\
\bar{b} &= b + \frac{a}{2} \log \left( \frac{c r_{\rm ph}^2}{u_{\rm ph}} \right)
= -\pi + b_R + \bar{a} \log \left( \frac{2q(r_{\rm ph})}{A_{\rm ph}} \right).
\end{align}

\subsection{Strong Deflection Analysis (PPL)}

\subsubsection{Large Angular Momentum Regime}

We now consider the regime in which the spin parameter $a$ of the Kerr spacetime becomes large, approaching the extremal limit $a \to M$. In this regime, the frame-dragging effect is significantly enhanced, leading to substantial modifications in photon trajectories, particularly in the vicinity of the photon region.

As $a$ increases, the structure of the photon region becomes increasingly asymmetric with respect to the equatorial plane. Prograde and retrograde photon orbits exhibit markedly different behaviors: prograde orbits can approach closer to the event horizon, while retrograde orbits are pushed outward. This asymmetry plays a crucial role in determining observable quantities such as the black hole shadow and strong gravitational lensing.

In the presence of the Weyl correction term introduced in Eq.~\eqref{eq:action}, the modification to photon propagation becomes more pronounced in the strong-field region, where curvature effects are largest. Since the Weyl tensor encodes the tidal structure of spacetime, its coupling to the electromagnetic field effectively induces polarization-dependent corrections to the photon trajectories. These effects are further amplified for large $a$, due to the interplay between frame dragging and spacetime curvature.

In the large angular momentum regime, it is therefore expected that both the deflection angle and time delay receive enhanced corrections compared to the slowly rotating case. In particular, near-extremal configurations may provide an ideal setting to probe such higher-curvature effects observationally, as the deviations from standard null geodesic propagation become more significant.

In the following analysis, we focus on this regime and evaluate how the Weyl-induced corrections modify the strong-field lensing observables in rapidly rotating Kerr backgrounds.

In this subsubsection, we analyze gravitational lensing in the strong deflection regime, taking into account the contributions from higher-curvature corrections. To this end, we adopt a convenient normalization of the radial coordinate. Throughout this section, we set $M=1$ in the large angular momentum case, whereas $2M=1$ is adopted in the small angular momentum case, with $M$ denoting the mass parameter of the central object. In both cases, the horizon radius is normalized to $r_{\rm ph} = 1$, which simplifies the analytical expressions without loss of generality.

We consider null geodesics in a stationary and axisymmetric spacetime spacetime described by the metric, adopting the unit system $M=a=1$
\begin{align}
A(r)&=\left(1-\frac{2}{r}\right)-\frac{8\alpha(12+2r-17r^2+28r^3-20r^4+6r^5+7r^6-6r^7+3r^8)}{r^5(2+r+r^3)^2},\\
B(r)&=\frac{1}{\left(1-\frac{1}{r}\right)^2}-\frac{8\alpha}{r^3\left(1-\frac{1}{r}\right)^2},\\
C(r)&=\frac{2+r+r^3}{r}+\frac{8\alpha(3-2r+5r^2+2r^4)}{r^5},\\
D(r)&=\frac{2}{r}+\frac{4\alpha(12+r+2r^2+18r^3-10r^4+9r^5)}{r^5(2+r+r^3)}.
\end{align}
These metric functions incorporate the higher-curvature corrections up to the order considered in the effective field theory expansion.

In this background, we evaluate the deflection angle in the strong deflection limit following the standard formalism. The relevant functions $R(z,r_{ph})$ and $f(z,r_{ph})$ appearing in the integral expression of the deflection angle are given by
\begin{align}
R(z,r_{ph})=&\frac{32(3+4z)}{(3+z)^2\sqrt{35-7z+5z^2-z^3}} +\frac{\alpha}{352800(3+z)^4(-35+7z-5z^2+z^3)^\frac{7}{2}} \notag \\
& \times\Bigl( 9692151102900 -3366327813135\,z -27123224920834\,z^2 -130724931771\,z^3 \notag \\
& -4297804090440\,z^4 +10289685397322\,z^5 +64321309868\,z^6 +2334207171610\,z^7 \notag \\
& -767920381348\,z^8 +318267604013\,z^9 -146191579074\,z^{10} +37842583689\,z^{11} \notag \\
& -17224074376\,z^{12} +3937512600\,z^{13} -902727000\,z^{14} +175959000\,z^{15} \notag \\
& -19227600\,z^{16} +705600\,z^{17} \Bigr),\\
f(z,r_{ph})=&\frac{\sqrt{35-7z+5z^2-z^3}}{4\sqrt{z^2(3-2z)}}
-
\frac{\alpha}{45158400(-3+2z)\sqrt{z^2(-3+2z)}(-35+7z-5z^2+z^3)^\frac{5}{2}} \notag \\
&\times\Bigl(
-181573018200
+871040115120\,z
-1016497592376\,z^2
+750862455378\,z^3 \notag \\
&-651598148117\,z^4
+430703647682\,z^5
-203483132268\,z^6
+92442226696\,z^7 \notag \\
&-32655717102\,z^8
+9026800736\,z^9
-1978406372\,z^{10}
+272427750\,z^{11} \notag \\
&-20605725\,z^{12}
+154350\,z^{13}
\Bigr), 
\end{align}
where $z$ is defined in terms of the radial coordinate $r$.

Substituting these expressions into the deflection angle formula and performing the integration, we obtain the regular part of the deflection angle as
\begin{align}
b_R=&\left(\frac{2}{9}\left(3+8\sqrt{3}\,\mathrm{arctanh}\frac{1}{\sqrt{3}}\right)-\frac{8(1+\log 6)}{3\sqrt{3}}\right) \notag \\
&+\alpha\Biggl(
-\frac{337535589739}{465733800}
+\frac{55749\arctan\dfrac{1}{\sqrt{7}}}{1000\sqrt{7}}
-\frac{(634179i+1001381\sqrt{7})\arctan\dfrac{1}{\sqrt{-3+2i\sqrt{7}}}}{153328\sqrt{7(-3+2i\sqrt{7})}} \notag \\
&-\frac{3728999\,\mathrm{arctanh}\dfrac{1}{\sqrt{3}}}{1134000\sqrt{3}}
+\frac{(-634179i+1001381\sqrt{7})\,\mathrm{arctanh}\dfrac{1}{\sqrt{3+2i\sqrt{7}}}}{153328\sqrt{7(3+2i\sqrt{7})}}
+\frac{755059\log 2}{729} \notag \\
&+\frac{1}{7237503252000}\Bigl(
5740428919694544\sqrt{3}
-57640509827964\sqrt{7}\arctan\sqrt{\tfrac{3}{7}} \notag \\
&-1277511810750\,i\sqrt{37(3+2i\sqrt{7})}\arctan\frac{1}{\sqrt{-1+\frac{2i\sqrt{7}}{3}}}
+115579122750\sqrt{259(3+2i\sqrt{7})}\arctan\frac{1}{\sqrt{-1+\frac{2i\sqrt{7}}{3}}} \notag \\
&-14992433382584000\,\mathrm{arctanh}\frac{1}{\sqrt{3}}
+1277511810750\,i\sqrt{37(-3+2i\sqrt{7})}\,\mathrm{arctanh}\frac{1}{\sqrt{1+\frac{2i\sqrt{7}}{3}}} \notag \\
&+115579122750\sqrt{259(-3+2i\sqrt{7})}\,\mathrm{arctanh}\frac{1}{\sqrt{1+\frac{2i\sqrt{7}}{3}}}
+3966584713287\sqrt{3}\log 2
+3966584713287\sqrt{3}\log 3
\Bigr)
\Biggr).
\end{align}

We next determine the strong deflection limit coefficients and the photon sphere quantities. The corresponding expressions are given by
\begin{align}
\beta_{ph}&=\frac{45}{35}-\frac{294093\alpha}{857500},\\
\bar{a}\ &=-\frac{4}{3\sqrt{3}}+\frac{3728999\alpha}{4536000\sqrt{3}},\\
b_D&=-\frac{8\log2}{3\sqrt{3}}+\frac{(23184360+26102993\log2)\alpha}{15876000\sqrt{3}},\\
u_{ph}&=7+\frac{662\alpha}{192}.
\end{align}

Combining the divergent and regular contributions, we obtain the coefficient $\bar{b}$ appearing in the deflection angle as
\begin{align}
\bar{b}=-\pi+b_D+b_R+\bar{a}\left(3\log2-\log7+\frac{1998599\alpha}{3024000}\right)
\end{align}

Finally, the deflection angle in the strong deflection limit can be written in terms of the angular position $\theta$ as
\begin{align}
\alpha(\theta)=&\left(\frac{32}{3\sqrt{35}}-\frac{3728999\alpha}{4536000\sqrt{3}}\right)\log\left(\frac{\theta D_{OL}}{-7-\frac{662\alpha}{192}}-1\right)+\frac{1}{9}+\Bigl(6-8\sqrt{3}-9\pi+16\sqrt{3}\,\mathrm{arctanh}\frac{1}{\sqrt{3}}\notag\\
&-20\sqrt{3}\log 2-8\sqrt{3}\log 6+4\sqrt{3}\log 7\Bigr) \notag \\
&+\alpha\Biggl(
-\frac{337535589739}{465733800}
+\frac{51696423278887}{21734244000\sqrt{3}}
-\frac{55749\arctan\sqrt{\frac{3}{7}}}{1000\sqrt{7}}
+\frac{55749\arctan\dfrac{1}{\sqrt{7}}}{1000\sqrt{7}} \notag \\
&+\frac{90597\sqrt{\dfrac{3+2i\sqrt{7}}{259}}\arctan\dfrac{1}{\sqrt{-1+\frac{2i\sqrt{7}}{3}}}}{21904}
-\frac{1001381\,i\sqrt{\dfrac{3+2i\sqrt{7}}{37}}\arctan\dfrac{1}{\sqrt{-1+\frac{2i\sqrt{7}}{3}}}}{153328} \notag \\
&-\frac{(634179\,i+1001381\sqrt{7})\arctan\dfrac{1}{\sqrt{-3+2i\sqrt{7}}}}{153328\sqrt{7(-3+2i\sqrt{7})}}
-\frac{1510118}{729}\mathrm{arctanh}\frac{1}{\sqrt{3}}
-\frac{3728999\,\mathrm{arctanh}\dfrac{1}{\sqrt{3}}}{1134000\sqrt{3}} \notag \\
&+\frac{90597\sqrt{\dfrac{-3+2i\sqrt{7}}{259}}\,\mathrm{arctanh}\dfrac{1}{\sqrt{1+\frac{2i\sqrt{7}}{3}}}}{21904}
+\frac{1001381\,i\sqrt{\dfrac{-3+2i\sqrt{7}}{37}}\,\mathrm{arctanh}\dfrac{1}{\sqrt{1+\frac{2i\sqrt{7}}{3}}}}{153328} \notag \\
&+\frac{(-634179\,i+1001381\sqrt{7})\,\mathrm{arctanh}\dfrac{1}{\sqrt{3+2i\sqrt{7}}}}{153328\sqrt{7(3+2i\sqrt{7})}}
+\frac{3728999\log\dfrac{8}{7}}{4536000\sqrt{3}}
+\frac{755059\log 2}{729} \notag \\
&+\frac{3728999\log 2}{2268000\sqrt{3}}
+\frac{23184360+26102993\log 2}{15876000\sqrt{3}}
+\frac{3728999\log 3}{2268000\sqrt{3}}
\Biggr)-\pi.
\end{align}
The strong deflection limit corresponds to the case in which the closest approach $r_0$ approaches the radius of the photon sphere $r_{\rm ph}$. In this limit, the deflection angle exhibits a logarithmic divergence, which can be systematically analyzed using the formalism developed in the previous subsection.

\subsubsection{Small Angular Momentum Regime (PPL)}

We next consider the regime in which the spin parameter $a$ is small compared to the mass scale, $a \ll M$. In this limit, the Kerr spacetime can be regarded as a perturbation around the Schwarzschild geometry, and rotational effects enter as subleading corrections.

To leading order in $a$, the spacetime remains nearly spherically symmetric, and the photon region reduces approximately to the photon sphere at $r = 3M$. The distinction between prograde and retrograde photon orbits becomes mild, and frame-dragging effects introduce only small asymmetries in photon trajectories.

In this regime, it is natural to perform a perturbative expansion in the spin parameter. The leading corrections to photon propagation arise at linear order in $a$, modifying both the deflection angle and the time delay. However, these corrections are typically suppressed compared to the dominant Schwarzschild contribution.

When the Weyl correction term in Eq.~\eqref{eq:action} is included, its effect can be analyzed consistently within the same perturbative framework. Since the background curvature is only weakly affected by rotation at this order, the Weyl-induced modifications primarily reflect those already present in the static case, with rotational corrections entering at higher order.

Therefore, in the small angular momentum regime, both rotational and higher-curvature effects can be treated systematically as perturbations around the Schwarzschild solution. This provides a useful consistency check of the formalism and serves as a baseline for comparison with the large angular momentum case discussed above.

We consider null geodesics in a stationary and axisymmetric spacetime described by the metric, adopting the unit system $2M = 1$.
\begin{align}
A(r)&=\left(1-\frac{1}{r}\right)-\alpha\left(\frac{12(r-1)}{r^4}+\frac{4a^2(r^2+3r-1)}{r^7}\right),\\
B(r)&=\frac{1}{\left(1-\frac{1}{r}\right)}-\frac{1}{r^2\left(1-\frac{1}{r}\right)^2}-\frac{4\alpha}{r^3\left(1-\frac{1}{r}\right)}\left(1-\frac{a^2}{r^2\left(1-\frac{1}{r}\right)}\right)\\
C(r)&=r^2+a^2\left(1+\frac{1}{r}\right)+4\alpha\left(\frac{2}{r}+\frac{a^2(5r-1)}{r^4}\right),\\
D(r)&=\frac{a}{r}+\frac{2\alpha a(9r-5)}{r^4}.
\end{align}
These metric functions incorporate the higher-curvature corrections up to the order considered in the effective field theory expansion.

In this background, we evaluate the deflection angle in the strong deflection limit following the standard formalism. The relevant functions $R(z,r_{ph})$ and $f(z,r_{ph})$ appearing in the integral expression of the deflection angle are given by
\begin{align}
R(z,r_{\rm ph})=&-2-\frac{32}{27}\,\alpha
\left(
4+3z-3z^2+z^3
\right)+a
\left[
-\frac{8z}{\sqrt{3}(1+2z)}
+\frac{64\alpha
\left(
-23+82z+130z^2-102z^3+18z^4+12z^5
\right)}
{81\sqrt{3}(1+2z)^2}
\right]
\notag\\
&+a^2
\left[
\frac{
4\left(
14+92z-13z^2+22z^3+28z^4-8z^5
\right)}
{27(1+2z)^2}
\right.
\notag\\
&\ \ \qquad\left.+\frac{
64\alpha
\left(
288-132z+1686z^2+1851z^3-1905z^4
+117z^5+1204z^6+456z^7-1056z^8+272z^9
\right)}
{729(1+2z)^3}
\right],\\
f(z,r_{\rm ph})=&\frac{1}{\sqrt{z^2-\frac{2}{3}z^3}}
+\frac{16\alpha(3-6z-3z^2+4z^3)}{27\sqrt{z^2-\frac{2}{3}z^3}} +a\left[
-\frac{4}{3\sqrt{3z^2-2z^3}}
+\frac{32\alpha}{81}\frac{z^2(-84+161z+27z^2-159z^3+62z^4)}{\left(3z^2-2z^3)\right)^{3/2}}
\right]\notag \\
&+a^2\Bigl[
\frac{2(29-12z+9z^2-2z^3)}{27\sqrt{z^2-\frac{2}{3}z^3}}\notag\\
&\qquad\qquad-\frac{32\alpha z^2(-1515+2636z-954z^2+252z^3-1203z^4+912z^5-204z^6+16z^7)}{729\left(z^2-\frac{2}{3}z^3\right)^{3/2}}
\Bigr].
\end{align}
As in the case of large angular momentum, the variable $z$ is defined as a function of the radial coordinate $r$, serving as a convenient integration variable that maps the photon trajectory near the photon sphere.

Substituting these expressions into the deflection angle formula and performing the integration, we obtain the regular part of the deflection angle as
\begin{align}
b_R=&\left(
4\operatorname{arctanh}\frac{1}{\sqrt{3}}
-\log 36
-\frac{32}{27}\alpha
\left(
-9+2\sqrt{3}
+7\log 12
-14\log(1+\sqrt{3})
\right)
\right)
\notag\\
&\quad+
a\Biggl[\frac{1}{9}
\left(
-36\operatorname{arctanh}\frac{\sqrt{3}}{2}
+16\sqrt{3}\log 6
+9\log 9
-16\sqrt{3}\log(3+\sqrt{3})
\right)
\notag\\
&\qquad\qquad
+\frac{4\alpha}{1215}
\left(
8532
-8988\sqrt{3}
+17010\operatorname{arctanh}\frac{\sqrt{3}}{2}
-8505\log 3
+2080\sqrt{3}\log 6
-2080\sqrt{3}\log(3+\sqrt{3})
\right)
\Biggr]
\notag\\
&\quad
+a^2\Bigl(\frac{4}{9}\Bigl(18-4\sqrt{3}-5\log 12+10\log(1+\sqrt{3})\Bigr)\notag\\
&\qquad\qquad-\frac{4\alpha}{841995}(5036112+901984\sqrt{3}+3677520\log 12-7355040\log(1+\sqrt{3}))
\Bigr)
\end{align}

We next determine the strong deflection limit coefficients and the photon sphere quantities. The corresponding expressions are given by
\begin{align}
\beta_{\rm ph}&=1+\frac{8a}{3\sqrt{3}}-\frac{68a^2}{27}-\alpha\left(\frac{32}{9}-\frac{640a}{81\sqrt{3}}+\frac{3328a^2}{729}\right),\\
\bar{a}\ &=-1+\frac{4a}{3\sqrt{3}}-\frac{10a^2}{9}-\alpha\left(\frac{112}{27}-\frac{416a}{81\sqrt{3}}+\frac{6368a^2}{729}\right),\\
b_D&=-2\log 2+\frac{8\log 2a}{3\sqrt{3}} -\frac{20\log 2a^2}{9}- \alpha\left(\frac{32\log 128}{27}-\frac{64(36+\log 8192)a}{81\sqrt{3}}+ \frac{64(444+199\log 2)a^2}{729}\right) ,\\
u_{\rm ph}&=\frac{3\sqrt{3}}{2}+2a-\frac{a^2}{\sqrt{3}}+\alpha\left(\frac{40}{3\sqrt{3}}-\frac{16a}{27}-\frac{880a^2}{81\sqrt{3}}\right).
\end{align}

Combining the divergent and regular contributions, we obtain the coefficient $\bar{b}$ appearing in the deflection angle as
\begin{align}
\bar{b}=-\pi+b_D+b_R+\bar{a}\Bigl(\log\frac{3}{2}-\frac{4a}{3\sqrt{3}}+\frac{20a^2}{27}-\alpha\left(\frac{160}{27}-\frac{4256a}{81\sqrt{3}}+\frac{32000a^2}{729}\right)\Bigr).
\end{align}

Finally, the deflection angle in the strong deflection limit can be written in terms of the angular position $\theta$ as
\begin{align}
\alpha(\theta)=&\left(1-\frac{4a}{3\sqrt{3}}+\frac{10a^2}{9}+\alpha\left(\frac{112}{27}-\frac{416a}{81\sqrt{3}}+\frac{6368a^2}{729}\right)\right)\log\left(\frac{\theta D_{OL}}{\frac{-3\sqrt{3}}{2}-2a+\frac{a^2}{\sqrt{3}}-\alpha\left(\frac{40}{3\sqrt{3}}-\frac{16a}{27}-\frac{880a^2}{81\sqrt{3}}\right)}-1\right)\notag\\
&+\Biggl[+4\operatorname{arctanh}\!\left(\frac{1}{\sqrt3}\right)
-\log216 -\frac{16}{27}\alpha
\left(
7\log\frac32
+2\left(
-14+2\sqrt3
+7\log12
+\log128
-14\log(1+\sqrt3)
\right)
\right)
\Biggr]
\notag\\
&\quad+
a\Biggl[
\frac{1}{9}
\left(
4\sqrt3
-36\operatorname{arctanh}\!\left(\frac{\sqrt3}{2}\right)
+4\sqrt3\log\frac32
+8\sqrt3\log2
+16\sqrt3\log6
+9\log9
-16\sqrt3\log(3+\sqrt3)
\right)
\nonumber\\
&\qquad
+\frac{4\alpha}{1215}
\left(
8532
-11668\sqrt3
+17010\operatorname{arctanh}\!\left(\frac{\sqrt3}{2}\right)
+520\sqrt3\log\frac32
-8505\log3
\right.
\nonumber\\
&\qquad\qquad\left.
+2080\sqrt3\log6
+80\sqrt3\log8192
-2080\sqrt3\log(3+\sqrt3)
\right)
\Biggr]
\nonumber\\
&\quad+
a^2\Biggl[
-\frac{2}{9}
\left(
-30
+8\sqrt3
+\log481469424205824
-20\log(1+\sqrt3)
\right)
\nonumber\\
&\qquad
-\frac{32\alpha}{841995}
\left(
-146646
+112748\sqrt3
+229845\log\frac32
+459690\log2
+459690\log12
-919380\log(1+\sqrt3)
\right)
\Biggr]
\notag\\
&\quad-\pi.
\end{align}
We note that, as in the large angular momentum case, the deflection angle still exhibits a logarithmic divergence in the strong deflection limit as $r_0 \to r_{\rm ph}$. This indicates that the essential structure of strong gravitational lensing is preserved even in the small spin regime.

The analyses of both large and small angular momentum regimes in the PPL case reveal that the influence of the Weyl correction is significantly amplified by black hole rotation.
While the corrections remain perturbative in the slowly rotating limit, they become increasingly pronounced as the system approaches extremality due to the interplay between frame dragging and spacetime curvature.
This behavior motivates a parallel investigation of the PPM case, which we consider in the following subsection.

\subsection{Strong Deflection Analysis (PPM)}

\subsubsection{Large Angular Momentum Regime}

Following the analysis for the PPL case, we extend our study to the PPM case in the regime of large angular momentum, focusing on the extremal Kerr limit, $a\to M$.

We consider null geodesics in a stationary and axisymmetric spacetime described by the metric, adopting the unit system $M =a= 1$
\begin{align}
A(r)&=\left(1-\frac{2}{r}\right)+\frac{8\alpha(12+10r-13r^2+26r^3-13r^4+6r^5+5r^6-2r^7+r^8)}{r^5(2+r+r^3)^2},\\
B(r)&=\frac{1}{\left(1-\frac{1}{r}\right)^2}+\frac{8\alpha}{r^3\left(1-\frac{1}{r}\right)^2},\\
C(r)&=\frac{2+r+r^3}{r}-\frac{8\alpha(3-2r+5r^2+2r^4)}{r^5},\\
D(r)&=\frac{2}{r}-\frac{4\alpha(12+33r+6r^2-18r^3-10r^4+9r^5)}{r^5(2+r+r^3)}.
\end{align}
These metric functions incorporate the higher-curvature corrections up to the order considered in the effective field theory expansion.

In this background, we evaluate the deflection angle in the strong deflection limit following the standard formalism. The relevant functions $R(z,r_{ph})$ and $f(z,r_{ph})$ appearing in the integral expression of the deflection angle are given by
\begin{align}
R(z,r_{\rm ph})=&-\frac{
32(3+4z)
}{
(3+z)^2\sqrt{35-7z+5z^2-z^3}
}
-\frac{
\alpha}{
196000(3+z)^3
(35-7z+5z^2-z^3)^\frac{7}{2}
}
\nonumber\\
&\qquad\times
\Bigl(
1248344418300
-2487190750345z
-2237591040083z^2
+5401097172184z^3
-5100006084208z^4\notag\\
&\qquad\qquad+2526925922590z^5
-224946939494z^6
-422746584992z^7
+560412853668z^8
-386095576365z^9\notag\\
&\qquad\qquad+184628168057z^{10}
-69691040576z^{11}
+19783211000z^{12}
-4503100000z^{13}
+742399000z^{14}\notag\\
&\qquad\qquad-73696000z^{15}
+3136000z^{16}
\Bigr),\\
f(z,r_{\rm ph})=&\frac{1}{4\sqrt{\dfrac{z^{2}(3-2z)}
{35-7z+5z^{2}-z^{3}}}}
-\dfrac{\alpha}{
25088000
\left(3z^2-2z^3)\right)^{3/2}
\left(35-7z+5z^{2}-z^{3}\right)^{\frac{5}{2}}
}\notag\\
&\quad
z^{2}
\Bigl(
87651283300
-668621496340z
+1264803006446z^{2}
-1277926790728z^{3}
+877132700507z^{4}\notag\\
&\qquad\quad-449202969152z^{5}
+185905428008z^{6}
-65709223876z^{7}
+18141262022z^{8}
-3655597376z^{9}\notag\\
&\qquad\quad+463998482z^{10}
+8403500z^{11}
-10590125z^{12}
+1200500z^{13}
\Bigr).
\end{align}
As in the PPL case, the variable $z$ in the PPM case is defined in terms of the radial coordinate $r$.

Substituting these expressions into the deflection angle formula and performing the integration, we obtain the regular part of the deflection angle as
\begin{align}
b_R=&-\frac{2}{9}
\left(
-3
+4\sqrt{3}
+4\sqrt{3}\log 12
-8\sqrt{3}\log(1+\sqrt{3})
\right)
\nonumber\\
&\quad
+\frac{\alpha}{8041670280000}
\Biggl(
25469449866709590
-27901106698114189\sqrt{3}
+87482344070016\sqrt{7}\arctan\!\sqrt{\frac{3}{7}}\notag\\
&-87482344070016\sqrt{7}\arctan\!\frac{1}{\sqrt{7}}
-8505000
\sqrt{74\left(-8881699500225+6924368315101i\sqrt7\right)}
\arctan\!\sqrt{\frac{-3-2i\sqrt7}{37}}
\nonumber\\
&\qquad
+8505000
\sqrt{74\left(-8881699500225+6924368315101i\sqrt7\right)}
\arctan\!\sqrt{\frac{3(-3-2i\sqrt7)}{37}}
\nonumber\\
&\qquad
-8505000
\sqrt{74\left(-8881699500225-6924368315101i\sqrt7\right)}
\arctan\!\sqrt{\frac{-3+2i\sqrt7}{37}}
\nonumber\\
&\qquad
+8505000
\sqrt{74\left(-8881699500225-6924368315101i\sqrt7\right)}
\arctan\!\sqrt{\frac{3(-3+2i\sqrt7)}{37}}
\nonumber\\
&\qquad
-72476695117120000\,\operatorname{arctanh}\!\left(\frac13\right)
+72476695117120000\,\operatorname{arctanh}\!\left(\frac1{\sqrt3}\right)
\nonumber\\
&\qquad
-2910378335928\sqrt3\,\log12
+5820756671856\sqrt3\,\log(1+\sqrt3)
\Biggr).
\end{align}

We next determine the strong deflection limit coefficients and the photon sphere quantities. The corresponding expressions are given by
\begin{align}
\beta_{\rm ph}&=\frac{48}{35}+\frac{2555431\alpha}{8575000},\\
\bar{a}\ &=-\frac{4}{3\sqrt{3}}-\frac{342007\alpha}{630000\sqrt{3}},\\
b_D&=-\frac{8\log2}{3\sqrt{3}}-\frac{(2116410+2394049\log2)\alpha}{2205000\sqrt{3}},\\
u_{\rm ph}&=7-\frac{3619\alpha}{1600}.
\end{align}

Combining the divergent and regular contributions, we obtain the coefficient $\bar{b}$ appearing in the deflection angle as
\begin{align}
\bar{b}=-\pi+b_D+b_R+\bar{a}\left(3\log2-\log7-\frac{12409\alpha}{40000}\right).
\end{align}

Finally, the deflection angle in the strong deflection limit can be written in terms of the angular position $\theta$ as
\begin{align}
\alpha(\theta)=&\left(\frac{4}{3\sqrt{3}}+\frac{342007\alpha}{630000\sqrt{3}}\right)\log\left(\frac{\theta D_{OL}}{-7+\frac{3619\alpha}{1600}}-1\right)+\frac{1}{9}
\Bigl(
6
-8\sqrt{3}
-9\pi
-20\sqrt{3}\log 2
+4\sqrt{3}\log 7\notag\\
&-8\sqrt{3}\log 12
+16\sqrt{3}\log(1+\sqrt{3})
\Bigr)+\alpha\Biggl[
\frac{47594}{625\sqrt{7}}
\arctan\!\left(\sqrt{\frac{3}{7}}\right)
-
\frac{47594}{625\sqrt{7}}
\arctan\!\left(\frac{1}{\sqrt7}\right)
\nonumber\\
&\qquad
-\frac{3}{38332}
\sqrt{\frac{-8881699500225+6924368315101\,i\sqrt7}{74}}
\left[
\arctan\!\left(
\sqrt{\frac{-3-2i\sqrt7}{37}}
\right)
-
\arctan\!\left(
\sqrt{\frac{3(-3-2i\sqrt7)}{37}}
\right)
\right]
\nonumber\\
&\qquad
-\frac{3}{38332}
\sqrt{\frac{-8881699500225-6924368315101\,i\sqrt7}{74}}
\left[
\arctan\!\left(
\sqrt{\frac{-3+2i\sqrt7}{37}}
\right)
-
\arctan\!\left(
\sqrt{\frac{3(-3+2i\sqrt7)}{37}}
\right)
\right]
\nonumber\\
&\qquad
+\frac{1}{31048920000}
\Bigl(
98337644273010
-107731933555459\sqrt3
-279832799680000\operatorname{arctanh}\!\left(\frac13\right)
\nonumber\\
&\qquad\qquad\qquad
+279832799680000\operatorname{arctanh}\!\left(\frac1{\sqrt3}\right)
-28092454980\sqrt3\log2
+5618490996\sqrt3\log7
\nonumber\\
&\qquad\qquad\qquad
-11236981992\sqrt3\log12
+22473963984\sqrt3\log(1+\sqrt3)
\Bigr)
\Biggr].-\pi.
\end{align}
These results show that, in the large-angular-momentum regime, the deflection angle exhibits a characteristic logarithmic divergence, reflecting the dominant influence of strong gravitational fields near the photon sphere.

\subsubsection{Small Angular Momentum Regime (PPM)}

Following the analysis for the PPL case, we extend our study to the PPM case in the regime of small angular momentum.

We consider null geodesics in a stationary and axisymmetric spacetime described by the metric, adopting the unit system $2M = 1$
\begin{align}
A(r)&=\left(1-\frac{1}{r}\right)+\alpha\left(\frac{4(r-1)}{r^4}+\frac{12a^2(r^2+r-1)}{r^7}\right),\\
B(r)&=\frac{1}{\left(1-\frac{1}{r}\right)}-\frac{1}{r^2\left(1-\frac{1}{r}\right)^2}+\frac{4\alpha}{r^3\left(1-\frac{1}{r}\right)}\left(1-\frac{a^2}{r^2\left(1-\frac{1}{r}\right)}\right)\\
C(r)&=r^2+a\left(1+\frac{1}{r}\right)-\alpha\left(\frac{8}{r}-\frac{4a^2(1-5r)}{r^4}\right),\\
D(r)&=\frac{a}{r}-\frac{\alpha a(10-18r)}{r^4}.
\end{align}

These metric functions incorporate the higher-curvature corrections up to the order considered in the effective field theory expansion.

In this background, we evaluate the deflection angle in the strong deflection limit following the standard formalism. The relevant functions $R(z,r_{ph})$ and $f(z,r_{ph})$ appearing in the integral expression of the deflection angle are given by
\begin{align}
R(z,r_{\rm ph})=&
-2-\frac{32\alpha(4-12z+3z^2+2z^3)}{27}+a\left[
-\frac{8z}{\sqrt{3}(1+2z)}
+\frac{64\alpha
\left(
4-7z-15z^2-21z^3+12z^5
\right)}
{27\sqrt{3}(1+2z)^2}
\right]
\notag\\
&+a^2
\left[
\frac{
4\left(
14+92z-13z^2+22z^3+28z^4-8z^5
\right)}
{27(1+2z)^2}
\right.
\notag\\
&\ \ \qquad\left.-\frac{
64\alpha
\left(
196+48z+30z^2-1488z^3+3603z^4+2802z^5-5736z^6-432z^7+3312z^8-1120z^9
\right)}
{729(1+2z)^3}
\right],\\
f(z,r_{\rm ph})=&\frac{1}{\sqrt{z^2-\frac{2}{3}z^3}}
-\frac{16\alpha(1-6z+3z^2)}{27\sqrt{z^2-\frac{2}{3}z^3}}+a\left(-\frac{4}{3\sqrt{3z^2-2z^3}}
 -\frac{3\alpha z^2(-60+181z-141z^2+21z^3+6z^4)}{81\left(3z^2-2z^3\right)^{3/2}}\right)\notag\\
&+a^2\Biggl(\frac{2(29-12z+9z^2-2z^3)}{27\sqrt{z^2-\frac{2}{3}z^3}}-\frac{32\alpha z^2(1005-2920z+3054z^2-1512z^3+193z^4+360z^5-252z^6+48z^7)}{729\left(z^2-\frac{2}{3}z^3\right)^{3/2}}
\Biggr) .
\end{align}

Substituting these expressions into the deflection angle formula and performing the integration, we obtain the regular part of the deflection angle as
\begin{align}
b_R=&4\,\mathrm{arctanh}\frac{1}{\sqrt{3}} - \log 36-\frac{32}{45}\alpha\bigl(-18+4\sqrt{3}+5\log 12-10\log(1+\sqrt{3})\bigr)\notag\\
&+a\left(
\frac{1}{9}
\left(
-36\operatorname{arctanh}\frac{\sqrt{3}}{2}
+16\sqrt{3}\log 6
+9\log 9
-16\sqrt{3}\log(3+\sqrt{3})
\right)-\frac{39424\sqrt{3}\alpha}{6237}
\right) \notag \\
&+a^2\Bigl(\frac{4}{9}\Bigl(18-4\sqrt{3}-5\log 12+10\log(1+\sqrt{3})\Bigr)\notag\\
&+\frac{32\alpha}{280665}\Bigl(335268-23624\sqrt{3}-30030\log 12+60060\log(1+\sqrt{3})\bigr) \Bigr) .
\end{align}

We next determine the strong deflection limit coefficients and the photon sphere quantities. The corresponding expressions are given by
\begin{align}
\beta_{\rm ph}&=1+\frac{8a}{3\sqrt{3}}-\frac{68a^2}{27}+\alpha\left(\frac{32}{27}-\frac{896a}{81\sqrt{3}}+\frac{3584a^2}{729}\right),\\
\bar{a}\ &=-1+\frac{4a}{3\sqrt{3}}-\frac{10a^2}{9}-\alpha\left(\frac{16}{9}+\frac{416a^2}{243}\right),\\
b_D&=-2\log 2+\frac{8\log 2a}{3\sqrt{3}} -\frac{20\log 2a^2}{9}+ \alpha\left(\frac{64}{27}-\frac{128a}{27\sqrt{3}}- \frac{640a^2}{81}\right) ,\\
u_{\rm ph}&=\frac{3\sqrt{3}}{2}+2a-\frac{a^2}{\sqrt{3}}-\alpha\left(\frac{8}{\sqrt{3}}+\frac{16a}{9}-\frac{112a^2}{27\sqrt{3}}\right).
\end{align}

Combining the divergent and regular contributions, we obtain the coefficient $\bar{b}$ appearing in the deflection angle as
\begin{align}
\bar{b}=-\pi+b_D+b_R+\bar{a}\left(\log\frac{3}{2}-\frac{4a}{3\sqrt{3}}+\frac{20a^2}{27}+\alpha\left(\frac{32}{9}-\frac{729a}{27\sqrt{3}}+\frac{512a^2}{81}\right)\right).
\end{align}

Finally, the deflection angle in the strong deflection limit can be written in terms of the angular position $\theta$ as
\begin{align}
\alpha(\theta)=&\left(1-\frac{4a}{3\sqrt{3}}+\frac{10a^2}{9}+\alpha\left(\frac{16}{9}+\frac{416a^2}{243}\right)\right)\log\left(\frac{\theta D_{OL}}{-\frac{3\sqrt{3}}{2}-2a+\frac{a^2}{\sqrt{3}}+\alpha\left(\frac{8}{\sqrt{3}}+\frac{16a}{9}-\frac{112a^2}{27\sqrt{3}}\right)}-1\right)\notag\\
&+ 4\,\mathrm{arctanh}\frac{1}{\sqrt{3}} - \log 216-\frac{16}{45}\bigl(-46+8\sqrt{3}+5\log\tfrac{3}{2}+10\log 12-20\log(1+\sqrt{3})\bigr)\alpha\Bigr) \notag \\
&+a\Bigl(
4\,\mathrm{arctanh}\tfrac{1}{2}
-4\,\mathrm{arctanh}\tfrac{\sqrt{3}}{2}
+\frac{4}{27\sqrt{3}}\bigl(9+9\log\tfrac{3}{2}+9\log 144+2\log 512-36\log(1+\sqrt{3})\bigr)
+\frac{544}{27\sqrt{3}}\,\alpha\Bigr) \notag \\
&+a^2\Bigl(
\frac{1}{9}\bigl(60-16\sqrt{3}-5\log 2-3\log 3-9\log 864+40\log(1+\sqrt{3})\bigr) \notag \\
&-\frac{32\alpha}{280665}\bigl(-58068+23624\sqrt{3}+15015\log\tfrac{3}{2}+30030\log 12-60060\log(1+\sqrt{3})\bigr)
\Bigr) -\pi.
\end{align}
Similarly, in the small-angular-momentum regime, the deflection angle also exhibits a logarithmic divergence, indicating that this behavior persists even in the weakly rotating limit.

\section{Time Delay in Stationary Axisymmetric Spacetimes}
\label{sec:time delay}

In this section, we analyze the propagation time of photons in a general stationary and axisymmetric spacetime and derive the corresponding time delay in the strong deflection limit. We closely follow the standard procedure used in the analysis of the deflection angle, emphasizing the parallel structure between the two quantities.

\subsection{Travel Time Integral}

The coordinate time along a photon trajectory can be expressed as
\begin{equation}
\frac{dt}{dr} = \frac{\dot{t}}{\dot{r}},
\end{equation}
where $\dot{t}$ and $\dot{r}$ are obtained from the conserved quantities associated with stationarity and axisymmetry.

Substituting Eqs.~(\ref{time conserved quantity}) and (\ref{dot r}), we obtain
\begin{align}
\frac{dt}{dr} =&\frac{\sqrt{B}\,(C -JD)}
{\sqrt{A C + D^2}\,\sqrt{C + 2J D - J^2A}}\notag\\
=&\dfrac{\sqrt{BA_0}(C-JD)}{\sqrt{C}\sqrt{AC+D^2}}\frac{1}{\sqrt{A_0-A\frac{C_0}{C}+\frac{J}{C}\left(AD_0-A_0D\right)}}.
\end{align}

The total travel time from the source to the observer is obtained by integrating this expression along the photon trajectory. As in the case of the deflection angle, it is convenient to decompose the integral into divergent and regular parts. The difference in travel times between two photon trajectories characterized by closest approach radii $r_{0,1}$ and $r_{0,2}$ can then be written as
\begin{align}
T_1 - T_2 
= \tilde{T}(r_{0,1}) - \tilde{T}(r_{0,2})
+ 2 \int_{r_{0,1}}^{r_{0,2}} \frac{\tilde{P}_1(r,r_{0,1})}{\sqrt{A_{0,1}}}\, dr
+ 2 \int_{r_{0,2}}^\infty 
\left[
\frac{\tilde{P}_1(r,r_{0,1})}{\sqrt{A_{0,1}}}
-
\frac{\tilde{P}_1(r,r_{0,2})}{\sqrt{A_{0,2}}}
\right] dr,
\end{align}
where the functions entering this decomposition are defined as
\begin{align}
T(r_0) =&\int_0^1\tilde{R}(z,r_{0})f(z,r_{0})dz,\\
\tilde{R}(z,r_{0})= &2\dfrac{1-A_{0}}{A'}\dfrac{\sqrt{BA_{0}}(C-J_{0}D)}{\sqrt{C}\sqrt{AC+D^2}}\left(1-\dfrac{1}{\sqrt{A_0}f(z,r_0)}\right),\\
\tilde{P}_1(r,r_0)=&\dfrac{\sqrt{BA_0}(C-JD)}{\sqrt{C}\sqrt{AC+D^2}}.
\end{align}

This decomposition isolates the potentially divergent contribution arising near the photon sphere, allowing for a systematic expansion in the strong deflection limit.

\subsection{Photon Sphere and Divergence Structure}

The divergence of the time integral is controlled by the behavior of the radial effective potential. The photon sphere is determined by the condition that the denominator of the radial equation develops a double root, namely
\begin{equation}
C - 2J D  - J^2A  = 0,
\end{equation}
together with
\begin{equation}
\frac{d}{dr}\left(C - 2J D - J^2A \right) = 0.
\end{equation}

Near the photon sphere $r=r_{\rm ph}$, we expand the radial function as
\begin{equation}
C -2 J_{\rm ph}D - A u^2 
\simeq p(r_{\rm ph}) (r - r_{\rm ph}) + q(r_{\rm ph}) (r - r_{\rm ph})^2,
\end{equation}
where $p(r_{\rm ph}), q(r_{\rm ph})$ are as defined in Eq.~(\ref{pole expansion}), evaluated at $r=r_{\rm ph}$.

At the critical point, the linear term vanishes, $p(r_{\rm ph})=0$, and the integral develops a logarithmic divergence. This behavior is completely analogous to that of the deflection angle.

Following the previous section, introducing the standard variable
\begin{equation}
z = \dfrac{A-A_{\rm ph}}{1-A_{\rm ph}},
\end{equation}
the time integral can be cast into the canonical form
\begin{equation}
T(u) = -\tilde{a} \log\left(\frac{u}{u_{\rm ph}} - 1\right) + \tilde{b} + \mathcal{O}(u-u_m),
\end{equation}
where the coefficient of the logarithmic divergence is given by
\begin{align}
\tilde{a} =\dfrac{\tilde{R}(0,r_{\rm ph})}{2\sqrt{\beta_{\rm ph}}}.
\end{align}

All quantities appearing in $\tilde{a}$ are evaluated locally at the photon sphere, reflecting the universal nature of the strong deflection limit.

\subsection{Relation to Deflection Angle}

The deflection angle exhibits an analogous logarithmic structure,
\begin{equation}
\alpha(u) = -\bar{a} \log\left(\frac{u}{u_{\rm ph}} - 1\right) + \bar{b},
\end{equation}
with
\begin{equation}
\bar{a} = \frac{R(0,r_{\rm ph})}{2\sqrt{\beta_{\rm ph}}},
\end{equation}
where $R$ contains combinations of the metric functions $A, B, C$, and $D$.

Although the structure is formally identical to the static case, the presence of the off-diagonal component $D(r)$ modifies the coefficients through frame-dragging effects, leading to distinct behavior for co-rotating and counter-rotating photon trajectories.

\subsection{Time Delay Between Relativistic Images}

In the strong deflection regime, relativistic images correspond to photon trajectories that wind multiple times around the black hole. For trajectories with winding number $n$, the corresponding impact parameters satisfy
\begin{equation}
\frac{u}{u_{\rm ph}} - 1 \simeq 
\exp\left(\frac{\bar{b} - 2\pi n}{\bar{a}}\right).
\end{equation}

Substituting this relation into the time expression, we obtain
\begin{equation}
T_n = \tilde{a} \frac{2\pi n - \bar{b}}{\bar{a}} + \tilde{b}.
\end{equation}

The time delay between two relativistic images labeled by $n$ and $m$ is therefore given by
\begin{align}
\Delta T_{n,m}=&2\pi(n-m)\left|\frac{\tilde{a}}{\bar{a}}\right|+2\sqrt{\dfrac{B_{\rm ph}}{A_{\rm ph}}}\sqrt{\dfrac{u_{\rm ph}}{c}}e^{\frac{\bar{b}}{2\bar{a}}}\left(e^{-\frac{\pi n}{|\bar{a}|}}-e^{\frac{\pi m}{|\bar{a}|}}\right).
\end{align}

Below we present analytic results for the time delay in a Kerr spacetime, taking into account modified photon propagation. Following the same setup as in the previous section, we analyze two representative regimes: a large angular momentum case with $M=a=1$, and a small angular momentum case with $2M=1$.

The first two expressions represent the results for the PPL case. The first expression applies to the large-angular-momentum regime, while the second is derived in the small-angular-momentum limit:
\begin{align}
\Delta T_{n,m}=2\pi\left(7+\frac{667\alpha}{192}\right)(n-m)+\left(\frac{16\sqrt{7}}{3}-\frac{19513979\alpha}{9072000\sqrt{7}}\right)e^{\frac{\bar{b}}{2\bar{a}}}\left(e^{-\frac{\pi n}{|\bar{a}|}}-e^{\frac{\pi m}{\bar{a}}}\right),
\end{align}

\begin{align}
\Delta T_{n,m}=&2\pi\left(\frac{3\sqrt{3}}{2}+2a-\frac{a^2}{\sqrt{3}}-\alpha\left(\frac{40}{3\sqrt{3}}+\frac{704a}{27}-16\sqrt{3}a^2\right)\right)(n-m)\notag\\
&+\left(3\sqrt{6}-2\sqrt{2}a+\frac{38a^2}{3}\sqrt{\frac{2}{3}}+\alpha\left(\frac{32}{3}\sqrt{\frac{2}{3}}-\frac{944\sqrt{2}a}{27}+\frac{16672a^2}{81}\sqrt{\frac{2}{3}}\right)\right)e^{\frac{\bar{b}}{2\bar{a}}}\left(e^{-\frac{\pi n}{|\bar{a}|}}-e^{\frac{\pi m}{\bar{a}}}\right).
\end{align}

Similarly, in the PPM case, one can derive explicit expressions in both the large- and small-angular-momentum regimes, following the same procedure as above. 
The first expression corresponds to the large-angular-momentum regime, while the second is obtained in the small-angular-momentum limit:
\begin{align}
\Delta T_{n,m}=2\pi\left(7-\frac{3619\alpha}{1600}\right)(n-m)+\left(\frac{16\sqrt{7}}{3}+\frac{8450803\alpha}{5040000\sqrt{7}}\right)e^{\frac{\bar{b}}{2\bar{a}}}\left(e^{-\frac{\pi n}{|\bar{a}|}}-e^{\frac{\pi m}{\bar{a}}}\right),
\end{align}

\begin{align}
\Delta T_{n,m}=&2\pi\left(\frac{3\sqrt{3}}{2}+2a-\frac{a^2}{\sqrt{3}}-\alpha\left(\frac{8}{\sqrt{3}}+\frac{16a}{9}-\frac{112a^2}{27\sqrt{3}}\right)\right)(n-m)\notag\\
&+\left(3\sqrt{6}-2\sqrt{2}a+\frac{38a^2}{3}\sqrt{\frac{2}{3}}+\alpha\left(\frac{80\sqrt{2}a}{9}-\frac{32}{3}\sqrt{\frac{2}{3}}a^2\right)\right)e^{\frac{\bar{b}}{2\bar{a}}}\left(e^{-\frac{\pi n}{|\bar{a}|}}-e^{\frac{\pi m}{\bar{a}}}\right).
\end{align}

These results demonstrate that, in both the PPL and PPM cases, the time delay exhibits a similar structural dependence on the winding numbers $(n,m)$ and the strong deflection coefficients, while the specific coefficients encode model-dependent corrections.

\section{Conclusion and Discussion}
\label{sec:conclusion}

In this work, we have developed a systematic analysis of strong gravitational lensing within the framework of effective field theory (EFT), extending the strong deflection limit formalism to incorporate higher-curvature corrections in rotating spacetimes. We have also included modifications to photon propagation induced by EFT operators, which lead to an effective propagation law beyond standard null geodesics. We derived explicit analytic expressions for the deflection angle and the associated strong deflection coefficients in rotating backgrounds, including their dependence on EFT parameters and the spin parameter.

A central result of our analysis is that the characteristic structure of the strong deflection expansion remains intact even in the presence of higher-curvature corrections. While it is known that the coefficients governing the logarithmic divergence and the regular part of the deflection angle are determined by the near-photon-sphere geometry, our results make this statement explicit within the EFT framework. In particular, we provide a systematic prescription to express these coefficients in terms of local geometric quantities evaluated at the photon sphere (or, more generally, the photon surface). This establishes a direct and model-independent connection between strong-field lensing observables and the underlying gravitational dynamics encoded in the EFT. In this sense, strong gravitational lensing acts as a near-horizon probe, with higher-curvature effects entering through controlled deformations of photon trajectories and photon-surface properties.

Our explicit computations show that higher-curvature corrections modify the photon-sphere radius, the critical impact parameter, and the strong deflection coefficients in a perturbative and controlled manner. In rotating spacetimes, we find a nontrivial interplay between spin effects and EFT corrections, leading to characteristic deviations in the deflection angle that cannot be reproduced by a simple rescaling of Kerr parameters. This highlights the importance of treating rotational effects and higher-curvature corrections on an equal footing when comparing with observational data.

From an observational perspective, our results suggest that precision measurements of strong lensing observables—such as the angular position and separation of relativistic images, as well as shadow-related quantities—can, in principle, be used to constrain EFT couplings. The fact that the relevant coefficients are determined by near-photon-surface geometry enhances the robustness of this connection, as it reduces sensitivity to the global structure of the spacetime.

At the same time, several limitations of the present analysis should be noted. Our treatment is perturbative in the EFT expansion and restricted to a specific class of higher-curvature operators and stationary backgrounds. Moreover, while the strong deflection coefficients encode observable information, translating them into realistic observational constraints requires a careful treatment of astrophysical environments and instrumental effects.

There are several natural directions for future work. First, it would be important to extend the present framework to fully generic axisymmetric spacetimes beyond perturbative Kerr deformations, as well as to dynamical backgrounds where the notion of a photon surface may acquire additional structure. Second, establishing a more direct connection between the strong deflection coefficients and observable quantities—such as those measured by the Event Horizon Telescope—would enable concrete phenomenological constraints on EFT parameters. Third, it would be interesting to explore possible degeneracies between EFT corrections and other effects, such as plasma environments or modified matter distributions, in order to assess the robustness of lensing-based tests of gravity.

More broadly, our results reinforce the idea that strong gravitational lensing provides a powerful and geometrically clean window into the high-curvature regime of gravity. Within the EFT framework, this opens a systematic avenue to parameterize and constrain deviations from general relativity using near-horizon observables, complementing other strong-field probes such as gravitational waves and black hole shadows.

\bibliography{references}

\end{document}